# A variable-coefficient model for decay of isotropic turbulence capturing effects of finite cascade time and Reynolds number

Rozie Zangeneh[1], Wenyuan Xue[1], Daniel Israel[2], and Ali Mani[1]
[1]*Department of Mechanical Engineering, Stanford University, Stanford, CA, USA*
[2]*Los Alamos National Laboratory, Ann Arbor, MI, USA*

We study isotropic turbulence decay in the context of the $k - \epsilon$ model, which solves the dissipation and kinetic energy equations. In modeling the dissipation equation, the coefficient $C_{\epsilon_2}$, suggested by Hanjalic and Launder [Journal of Fluid Mechanics, 1972] [1], is related to the temporal decay power-law by $n = 1/(C_{\epsilon_2} - 1)$ and is assumed to be a constant value. In this work, we perform high-fidelity numerical simulations to examine the mathematical terms responsible for the decay of isotropic turbulence, considering both scenarios of forced and decaying turbulence. Our data suggests that the instantaneous $C_{\epsilon_2}$ not only depends on the instantaneous Reynolds number but is also sensitive to the history of energy injection in turbulence. We attribute these observations to the finite time required for the cascade from energetic to dissipative scales. Considering data from both decaying and growing forced turbulence, we develop an evolution equation for $C_{\epsilon_2}$ with Reynolds-dependent coefficients. We demonstrate that this model accurately captures the time evolution of dissipation and kinetic energy over a wide range of Reynolds numbers under a wide range of forced and decay scenarios.

## 1 INTRODUCTION

Advancements in computing power have greatly broadened the accessibility of direct numerical simulations (DNS) for fundamental understanding of turbulent fluid flows. However, in most engineering applications, DNS is not yet feasible. At the same time, for the purpose of engineering analysis and design, often access to mean fields, such as ensemble-mean velocity data, is sufficient. In this context, the most straightforward approach would be to compute such quantities directly in the averaged space as done through the Reynolds-Averaged Navier Stokes (RANS) framework. There exists significant demand for improved RANS turbulence models that are informed by and can represent more comprehensive features of turbulence physics. Common RANS models offer a mathematical representation of closure quantities in the momentum equation in terms of solutions to additional differential equations that must be numerically evolved together with the mean momentum equation. Depending on the number of these additional equations, these models are referred to as one-equation model, two-equation model, etc [2]. The solution of these equations provides the length and/or time scales of the active part of the turbulent motion. The turbulence kinetic energy (TKE) equation is, without exception, employed in all two-equation turbulence models. A more general alternative to the TKE equation, which is employed in some models, is the Reynolds Stress Transport equation (RST). The RST equations include terms describing transport, generation, decay, and the inter-component energy transfer, which may be written as

$$\frac{d\overline{u_i u_j}}{dt} = P_{ij} + T_{ij} + \Pi_{ij} - \epsilon_{ij}, \tag{1}$$

where overbar, $\bar{\cdot}$ represents the ensemble average, and the velocity field is decomposed into a mean and fluctuation as $U_i + u_i$. On the left-hand side $d/dt$ represents the material derivative as $\partial/\partial t + U_k \partial/\partial x_k$. In the expression above, $P_{ij}$ is the Reynolds stress production equal to $P_{ij} = -\overline{u_i u_k}\frac{\partial U_j}{\partial x_k} - \overline{u_j u_k}\frac{\partial U_i}{\partial x_k} + \overline{u_i f_j + u_j f_i}$, where the second term augments Reynolds stress production only in the presence of a body force $f_i$, in the Navier-Stokes equation [3, 4]. $T_{ij} = \nu \frac{\partial^2 \overline{u_i u_j}}{\partial x_k x_k} - \frac{\partial \overline{u_i u_j u_k}}{\partial x_k} - \frac{1}{\rho}\left(\frac{\partial \overline{u_i p}}{\partial x_j} + \frac{\partial \overline{u_j p}}{\partial x_i}\right)$ is the transport term, $\Pi_{ij} = \frac{1}{\rho}\overline{p\left(\frac{\partial u_i}{\partial x_j} + \frac{\partial u_j}{\partial x_i}\right)}$ is the pressure-strain rate correlation and $\epsilon_{ij} = 2\nu \overline{\frac{\partial u_i}{\partial x_k}\frac{\partial u_j}{\partial x_k}}$ is the dissipation tensor. The TKE equation, mentioned earlier, is simply a scalar equation effectively representing the trace of the RST equation.

To allow the development of dimensionally consistent closures, models based on the TKE or RST equations need to be augmented by additional evolution equations that describe fields other than velocity moments. Given the importance of dissipative effects in zones of active turbulence, the turbulent kinetic energy dissipation equation is a popular choice for such an equation, which is also quantifiable from high-fidelity data. The exact transport equation for the kinetic energy dissipation rate ($\epsilon = \nu \overline{\frac{\partial u_i'}{\partial x_j} \frac{\partial u_i'}{\partial x_j}}$) in an incompressible flow can be derived from the momentum equation as follows (Mansour et al. [5]):

$$\frac{d\epsilon}{dt} = P^1 + P^2 + P^3 + P_\epsilon + F_\varepsilon + T + \Pi + D - Y, \tag{2}$$

where $P^1 = -2\nu \overline{\frac{\partial u_i'}{\partial x_j} \frac{\partial u_k'}{\partial x_j}} \frac{\partial U_k}{\partial x_i}$ is the mixed production, $P^2 = -2\nu \overline{\frac{\partial u_i'}{\partial x_j} \frac{\partial u_i'}{\partial x_k}} \frac{\partial U_k}{\partial x_j}$ is the production by mean velocity gradient, $P^3 = -2\nu \overline{u_k' \frac{\partial u_i'}{\partial x_j}} \frac{\partial^2 U_i}{\partial x_j \partial x_k}$ is the gradient production, $P_\epsilon = -2\nu \overline{\frac{\partial u_i}{\partial x_j} \frac{\partial u_k}{\partial x_j} \frac{\partial u_i}{\partial x_k}}$ is the turbulent production, $F_\varepsilon = 2\nu \overline{\frac{\partial u_i}{\partial x_j} \frac{\partial f_i}{\partial x_j}}$ is the production due to forcing, $T = -\nu \frac{\partial}{\partial x_k} \left( \overline{u_k \frac{\partial u_i}{\partial x_j} \frac{\partial u_i}{\partial x_j}} \right)$ and $\Pi = -\frac{2\nu}{\rho} \frac{\partial}{\partial x_k} \left( \overline{\frac{\partial u_k}{\partial x_j} \frac{\partial p}{\partial x_j}} \right)$ are turbulent and pressure transport terms, $D = \nu \frac{\partial^2 \epsilon}{\partial x_k \partial x_k}$ is the viscous diffusion and $Y = 2\nu^2 \overline{\left( \frac{\partial^2 u_i}{\partial x_j \partial x_k} \right) \left( \frac{\partial^2 u_i}{\partial x_j \partial x_k} \right)}$ is the viscous dissipation term.

For homogeneous and isotropic turbulence, which is the subject of this study, the transport terms are zero in both the RST and dissipation equations. Additionally, due to isotropy, these equations can be simplified to [6]

$$\frac{dk}{dt} = P - \epsilon, \tag{3}$$

$$\frac{d\epsilon}{dt} = P_\epsilon + F_\varepsilon - Y, \tag{4}$$

where $k$ is the turbulent kinetic energy, and is equal to half of the trace of the Reynolds stress tensor, and $P$ is half of the trace of the production tensor. In the absence of active forcing, where turbulence experiences a free decay, these equations are further simplified to

$$\frac{dk}{dt} = -\epsilon, \tag{5}$$

$$\frac{d\epsilon}{dt} = P_\epsilon - Y. \tag{6}$$

This limit of turbulence behavior, i.e., decaying isotropic turbulence, is often considered as the first building block of developing turbulence models. The TKE equation and the dissipation equation provide a standard framework for the construction and evaluation of such models, which would naturally lead to the $k - \epsilon$ family of models [7–11]. An extension of this family to models that accommodate anisotropy replaces the TKE equation with the full Reynolds stress equation [12–15]. Some of the other modeling families, such as $k - \omega$ [16–19] and the $k - L$ [20, 21] also use decaying isotropic turbulence as a key, often the first, building block of model construction. Typically, the model coefficients that govern the free-decay behavior are tuned first. Subsequent tuning steps, for example, in modeling transport terms, often via eddy diffusivity models, and models for near-wall behavior, are all influenced by this first step. As a result, an improved understanding and quantitative modeling of isotropic turbulence decay has implications relating to broad regimes of turbulence and beyond just homogeneous isotropic turbulence.

The advantage of the $k - \epsilon$ framework over other frameworks, such as $k - \omega$ and $k - L$, is that the second variable, $\epsilon$, is a measurable quantity, while variables $\omega$ and $L$ are merely auxiliary variables increasing the order and dimensional diversity of the model evolution equation. They are related to $\epsilon$ via scaling analysis as $L \sim k^{3/2}/\epsilon$, and $\omega \sim \epsilon/k$, but the community treats these as guiding scaling relations without committing to an exact definition of these variables. It should further be noted that even in the $k - \epsilon$ modeling family, the variable $\epsilon$ is often treated as a scale variable without committing it to be exactly equal to the defined dissipation rate. In this study, however, since we use DNS data, we quantify the exact $\epsilon$ to seek model improvement for predicting the decay behavior. Nevertheless, key quantities of interest, such as decay exponent, are insensitive to whether a model uses exact or some scaled form of $\epsilon$. With that in mind, we continue discussing closure models in the limit of homogeneous and isotropic turbulence. In this limit, Eq. 5 is already closed, and one should seek to obtain a mathematical model representing the right-hand side of Eq. 6.

As we shall see in Section 2.3, the terms $P_\epsilon$ and $Y$ diverge with increasing the Reynolds number. Dimensionally, that would be the limit of decreasing the kinematic viscosity while keeping the velocity and large eddy length fixed. However, it is well established that dissipation itself remains largely insensitive to the Reynolds number since it is primarily slave to the energy cascade from large scales. These two seemingly contrasting observations can be reconciled by the observation that the combined terms $P_\epsilon$ - $Y$, which we later call them together dissipation decay term, remain insensitive to the Reynolds number. In Section 2.3, we provide quantitative evidence of this observation. Therefore, it would be reasonable to attempt a closure model that first lumps these two terms and then expresses them in terms of an algebraic expression. This idea has historical precedent: following Rodi [22], Hanjalic and Launder [1] argued that the dissipation decay should be modeled through the combined term $P_\epsilon - Y$ rather than by treating $P_\epsilon$ and $Y$ separately. The DNS results presented later in Fig. 2 support this interpretation by showing that, although $P_\epsilon$ and $Y$ individually become large and Reynolds-number dependent, their combination remains much better behaved.

Using dimensional analysis and assuming insensitivity of the combined right-hand side of Eq. 6, one can arrive at the following model form for Eq. 6 [23].

$$\frac{d\epsilon}{dt} = -C_{\epsilon_2}\frac{\epsilon^2}{k}, \tag{7}$$

where $C_{\epsilon 2}$ is a tunable coefficient. For example, the standard $k-\epsilon$ model uses $C_{\epsilon 2} = 1.92$. An analytical solution to Eq. 6, combined with Eq. 5, results in the prediction of a power law decay of kinetic energy for homogeneous and isotropic turbulence.

$$k = k_0\left(\frac{t_0 + t}{t_0}\right)^{-n} \tag{8}$$

where $n$ is the decay exponent and equal to [24]

$$n = \frac{1}{C_{\epsilon 2} - 1}, \tag{9}$$

$k_0$ is the initial turbulent kinetic energy, and $t_0$ is a time scale, related to the initial kinetic energy and dissipation as $t_0 = nk_0/\epsilon_0$.

Despite the simplicity of decaying isotropic turbulence, no universal decay exponent has been confirmed throughout extensive computational and experimental works. As a result, several different outcomes have been reported in the literature. Notable experiments included those performed by Comte-Bellot and Corrsin [25, 26], finding a decay exponent of about $n \approx 1.25$, corresponding to $C_{\epsilon_2} \approx 1.8$. Values close to this have since been reported more recently in experiments, including those reported in Kang et al. [27] and Krogstad and Davidson [28]. DNS of decaying isotropic turbulence, however, have tended to report decay exponents in a broad range [29–31]. These studies are generally restricted to lower Reynolds numbers compared to the experiments. A recent survey of $n$ values from both experiments and simulations reveals that the decay of isotropic turbulence depends, presumably, on additional factors such as the instantaneous Reynolds number. Additionally, the results indicate that the time evolution of $n$ involves a transition, and a nearly constant value is obtained only after a sufficiently long time [32**? ? ?** , 33]. Thormann and Meneveau [34] discussed that initial conditions may linger for extended periods of time and qualitatively change the flow evolution. These observations imply that the decay rate $n$ depends on initial conditions, and the decay may not even follow a power-law, i.e., the coefficient $C_{\epsilon_2}$ may depend on time during the decay.

In this context, some models have been developed to account for the variability of $C_{\epsilon_2}$ due to the initial condition. One notable model, by Launder and Schiestel [35], recognizes that turbulent kinetic energy and rate of kinetic energy dissipation are dominated by large and small scales, respectively. Given this observation, the authors developed a multi-scale model by adding an evolution equation to track accounting of $k$ separately for large and small scales as well as an equation for the energy flux:

$$\frac{dk_L}{dt} = -f \tag{10a}$$

$$\frac{dk_S}{dt} = f - \epsilon \tag{10b}$$

$$\frac{df}{dt} = \frac{f}{k_L}(-C_{f_2} f) \tag{10c}$$

$$\frac{d\epsilon}{dt} = \frac{\epsilon}{k_S}(C_{\epsilon_1} f - C_{\epsilon_2}\epsilon) \tag{10d}$$

where TKE is broken into a large-scale part $k_L$, and a small-scale part $k_S$, and $f$ is the energy flux responsible for extracting TKE from large scales. This results in four equations and three coefficients. The model is intuitively understood as follows: the large-scale dynamics governed by $k_L$ and $f$ are modeled as in the traditional $k - \epsilon$ model and independently of the small scales. The small-scale dynamics for the dissipation equation are slaves to the large scale in a one-way coupled fashion by allowing a first-order damped equation for the evolution of kinetic energy dissipation. The small-scale kinetic energy equation is derived as an exact equation, given that total turbulent kinetic energy must be matched by total dissipation. A subsequent study by Klein et al. [36, 37] confirmed the expectation of possible improvements in predicting non-equilibrium flows using this multi-scale model. While this model is reasonably grounded on physical intuition, a challenge in its quantitative assessment and tuning is due to the fact that quantities $k_L$, $k_s$, and $f$ are not defined as measurable quantities. This issue limits the aforementioned advantage of the $k - \epsilon$ framework over other frameworks in terms of the possibility to directly probe the model's input parameters from data.

The original contribution of this work is the development of a model that allows variability of $C_{\epsilon_2}$ while maintaining a measurable input space. In doing so, we utilize model forms that result in a first-order damped behavior qualitatively similar to those envisioned in the multi-scale model of [35]. We do so by considering $C_{\epsilon_2}$ itself as a variable in the input space, governed by an evolution equation. By constructing a wide range of isotropic turbulence simulations, we directly measure the evolution of $C_{\epsilon_2}$ and capture its dynamics by a proposed model. Furthermore, our model not only accounts for the temporal evolution of $C_{\epsilon_2}$ due to initial conditions, but it also accounts for the effects of turbulence forcing as well as finite Reynolds number effects.

The rest of this report is organized as follows. Section II describes the methodology, including high-fidelity simulations and the formulation of the model. Simulation results are presented in Section III, followed by model tuning and model assessment. Finally, discussion and future work are presented in Section IV.

# 2 COMPUTATIONAL METHODOLOGY

## 2.1 Governing Equations

The data for tuning and testing models is generated by solving the following incompressible momentum equation on a uniform isotropic grid using a pseudospectral code adapted from the finite volume code of [38]

$$\frac{\partial u_i}{\partial t} + \frac{\partial u_j u_i}{\partial x_j} = -\frac{1}{\rho}\frac{\partial p}{\partial x_i} + \frac{\partial}{\partial x_j}\left(\nu \frac{\partial u_i}{\partial x_j}\right) + \Omega A_{ij}\tilde{u_j}, \tag{11}$$

in addition to the continuity equation, $\nabla \cdot \boldsymbol{u} = 0$. In the momentum equation, the last term is the forcing term, which may be adjusted to mimic different scenarios, including stationary force turbulence, decay, or growth of turbulence. The details on the forcing term are discussed below. For model development and tuning purposes, we conduct DNS to generate data of finite Reynolds numbers. For the purpose of testing the model, to represent the limit state of infinite Reynolds number turbulence, we perform large-eddy simulations (LES) by employing an eddy viscosity, $\nu_t$, instead of molecular viscosity, $\nu$, as done by Homan et al. [4]. This means all quantities in Eq. 11 are implicitly filtered, and the filter width is implied to be proportional to the grid size. In the context of LES to capture subgrid effects, we use a Smagorinsky-Lilly model, where $\nu_t = (C_s\Delta)^2|\overline{S}|$ [39]. In accordance with a constant-coefficient model, here $C_s = 0.2$ consistent with [40], $\Delta$ is the grid size, and $|\overline{S}| = \sqrt{2S_{ij}S_{ij}}$ is the magnitude of the resolved strain-rate tensor. Additionally, one needs to account for the Reynolds stress in subgrid scales; however, Homan et al. [4] instead performed a mesh convergence study and demonstrated small subgrid Reynolds Stress, given their resolution.

The last term in Eq. 11 is a linear forcing term that provides turbulent energy production or removal based on elements of the forcing matrix $A_{ij}$. This forcing technique is developed based on the works of [41, 42]. In this work, given our focus on isotropic turbulence, we chose $A_{ij} = A\delta_{ij}$, where $\delta_{ij}$ is the Kronecker delta. When simulating stationary turbulence, $\Omega$ is deployed as a time-varying controller that maintains the TKE at a prescribed level, using a proportional controller framework inspired by [3]. As discussed below, our forcing term is further modified to incorporate filtered velocity $\tilde{u_j} \neq u_j$ adopted by [4, 43] in order to maintain a domain-independent turbulence. The resulting production term in Eq. 1 is $P_{ij} = \overline{u_i\tilde{u}_k\Omega A_{jk}} + \overline{u_j\tilde{u}_k\Omega A_{ik}}$ as an alternative to the standard production $P_{ij} = -\overline{u_iu_k}\frac{\partial u_j}{\partial u_k} - \overline{u_ju_k}\frac{\partial u_i}{\partial u_k}$.

When simulating decaying turbulence, $\Omega$ is set to zero, while simulating growing turbulence, $\Omega A_{ij}$ is set as $\delta_{ij}A$ for generating training data. This implies that the forcing time scale is held fixed. However, in our testing stage, discussed in Section 3.2, we assess the performance of the developed model over a range of forcing time scales. To minimize the sensitivity of the results to domain size and orientation, we deployed a high-pass filter on the velocity field in the forcing term, $\tilde{u_i}$, similar to [43]. This filter, applied to the forced velocity field, smoothly varies from 0 at $\kappa = 2$ to 1 at $\kappa = 3$ using a cosine function. As a result, energy is only injected at wave numbers $\kappa > 2$. This strategy has been used extensively as a remedy to develop domain-independent forced turbulence [44–46]. It was argued that filtering is essential for maintaining a stationary and well-resolved energy spectrum, otherwise, any energy added at the low wavenumbers would accumulate since little dissipation is occurring at these scales, and therefore turbulence would be controlled by the domain size [4].

## 2.2 Numerical Setup

**Table 1. Simulation setup for stationary turbulence**

| $N$ | $\nu$ | $u_{rms}$ | $\epsilon$ | $L_I$ | $T_{eddy}$ | $Re_T$ |
|---|---|---|---|---|---|---|
| 128 | 0.011 | 1.00 | 2.86 | 0.34 | 0.35 | 78 |
| 192 | 0.0055 | 1.00 | 3.07 | 0.32 | 0.33 | 133 |
| 256 | 0.0042 | 1.00 | 3.12 | 0.29 | 0.33 | 180 |
| 384 | 0.0029 | 1.00 | 3.13 | 0.27 | 0.32 | 240 |
| 512 | 0.0020 | 1.00 | 3.10 | 0.28 | 0.32 | 382 |

The computational domain consists of a triply periodic box with nominal edge length $L = 2\pi$. Additionally, for simulations of stationary cases, the turbulence forcing term is controlled to produce TKE equal to 3/2, maintaining $u_{rms} = \sqrt{2k/3} = 1$. With these normalizations, in the presentation of results, time is nondimensionalized by $T = L/2\pi u_{rms}$. In our setup $A_{ij} = A\delta_{ij}$. The simulation outcome is not sensitive to the choice of nonzero amplitude $A$, since the controller $\Omega(t)$ would automatically adjust the magnitude of the forcing term to maintain the specified $u_{rms}$.

Various Reynolds numbers considered here are tabulated in Table 1. In the table, for each Reynolds number case, the number of mesh points is preselected, and the molecular viscosity is then chosen as the smallest value that still allows a fully resolved DNS on that grid. Specifically, the mesh was verified to be at most equal to twice the post-processed Kolmogorov length scale $\eta = (\frac{\nu^3}{\epsilon})^{1/4}$ [47] aligned with the study of Donzis et al. [48]. In our study, this resolution was found to be appropriate for obtaining closed budgets for both the Reynolds stress equation and the kinetic energy dissipation equation, as discussed below. Additionally, the table reports the post-processed $u_{rms}$, which is confirmed to be close to 1, given the controller. Furthermore, the post processed dissipation, $\epsilon$, and $T_{eddy} = u_{rms}^2/\epsilon$ is reported. For each case, the resulting turbulent Reynolds number $Re_T = k^2/\nu\epsilon$ is computed after post-processing the dissipation data. The integral length,$L_I$, which is the integral of autocorrelation in space normalized by $u_{rms}^2$, is reported in Table 1 for stationary turbulence. These values change while turbulence decays or as it grows. The maximum value of $L_I$ reaches 0.400 for the decay and 0.432 for the growth. Note that despite the variation, $L_I$ values remain small compared to the box size, namely $L_I < 2\pi/10$. To further establish confidence in mesh-independent results, we recompute the case with $\nu = 0.0055$ using a larger mesh size of 384. The results are shown in Table 2. It may be deduced that the post-processed values are very close for these two mesh sizes with the same viscosity. Therefore, we confirm that the results are converged with respect to mesh size.

**Table 2. Mesh convergence study for stationary forced simulations**

| $N$ | $\nu$ | $-Y$ | $P_\epsilon$ | $\epsilon$ | $C_\epsilon$ | $Re_T$ |
|---|---|---|---|---|---|---|
| 192 | 0.0055 | -28.18 | 21.76 | 3.07 | 1.02 | 133 |
| 384 | 0.0055 | -27.83 | 21.52 | 3.05 | 1.02 | 134 |

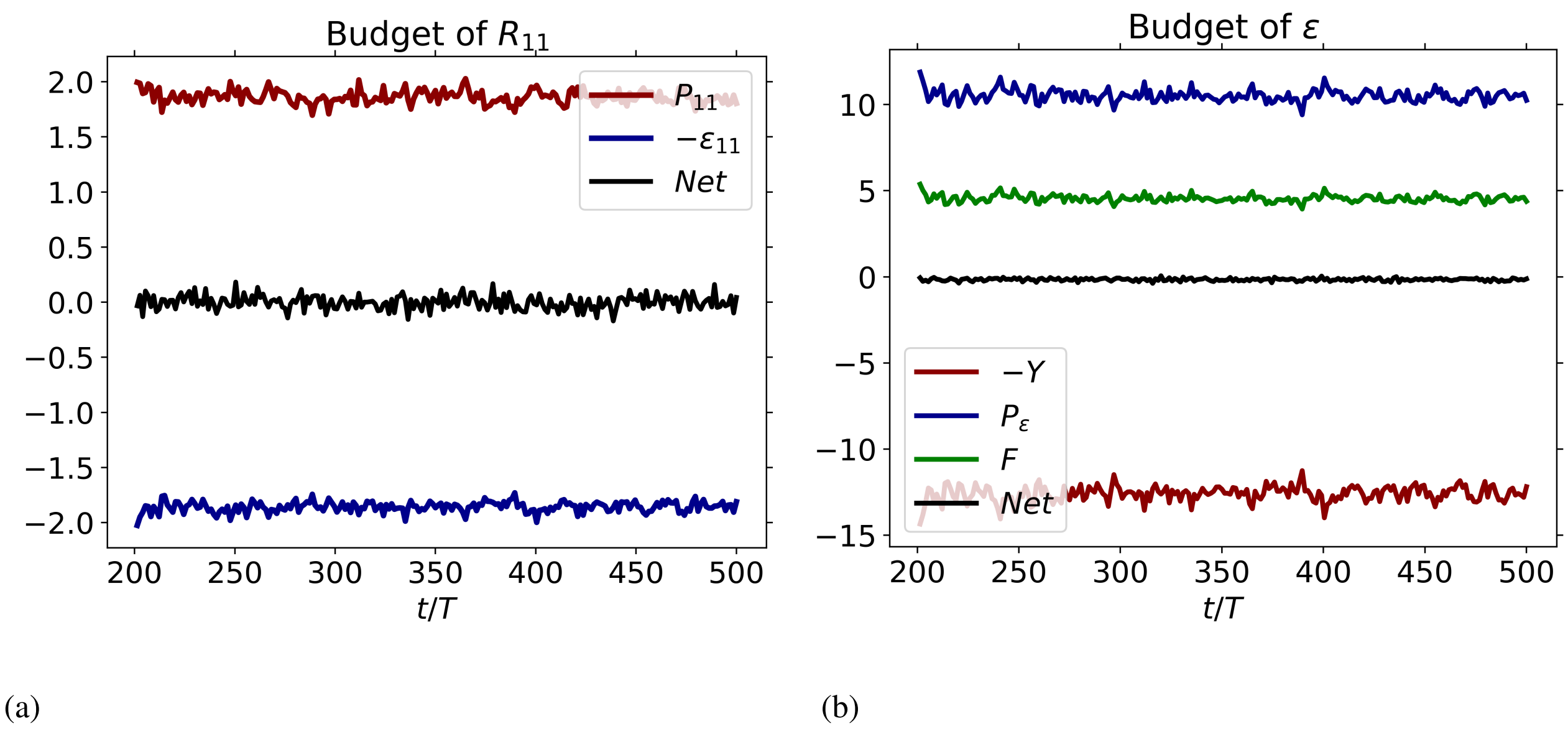

(a) (b)

**Figure 1. Examples of budgets for the Reynolds stress equation (a) and dissipation equation (b).**

The code uses Fourier spectral derivatives with explicit time advancement through a fourth-order Runge-Kutta method. Convective terms were dealiased using zero-padding with 3/2 rule. The closure of Reynolds stress budget and dissipation budget in the domain are confirmed, and examples are shown in Fig. 1a and Fig. 1b, respectively, for $N = 128$. Note that only the budget analysis of the x-component of Reynolds stress is shown here, since simulations are isotropic. Stationary forced simulations are run for $t/T = 400$ while the first $t/T = 100$ statistics are removed, and 150 ensembles with sampling intervals of $2T$ are used for statistical averaging of force turbulence to ensure independence of different ensembles.

## 2.3 Computational Results

We examined the decay term in the dissipation equation (Eq. 4), $P_\epsilon - Y$, for forced stationary turbulence as the Reynolds number varies. As plotted in Fig. 2, each term diverges with increasing Reynolds number, while the superposition of the two, which represents the decay term in the dissipation equation, remains convergent at higher Reynolds numbers. A leading-order analysis can explain the divergence of the individual terms with Reynolds number as follows. For example, taking the term $Y$ defined after Eq. 2, this term is proportional to the square of the second gradient of the velocity with the prefactor of the square of the molecular viscosity. Given that second derivatives are dominated by small scales, a scaling estimate for $Y$ can be written as

$$Y \sim \nu^2 u_\eta^2 \eta^{-4},$$

where $\eta$ and $u_\eta$ represent the Kolmogorov length and velocity scales, respectively. We next assume a sufficiently large Reynolds number and a Kolmogorov spectrum, and thereby estimate $\eta \sim l Re_T^{-3/4}$ [24, 47], where $l$ is the large eddy length, which is fixed in our setting by the filter length of the forcing term. We also use the relations $\nu = \eta u_\eta$, and $u_\eta \sim u_l (\eta/l)^{1/3}$, where $u_l$ is the large eddy velocity controlled by $u_{rms}$. Combining these substitutions, one can obtain the following leading-order estimate for $Y$ versus the Reynolds number.

$$Y \sim u_l^4 / l^2 Re_T^{1/2}. \tag{12}$$

Given this leading-order scaling, we hypothesize the dependence of $Y$ on $Re_T$, keeping large scale length and velocity fixed, to be in the form of an expansion involving $Re_T^{1/2}$ as $Y = y_1 Re_T^{1/2} + y_0 + y_{-1} Re_T^{-1/2} + y_{-2} Re_T^{-1} + \ldots$. The green solid line in Fig. 2 shows a fit to the data of $Y(Re_T)$ normalized by $\epsilon^2/k$ using only two terms of this expansion. One can derive a similar leading term and hypothesize a similar expansion for $P_\epsilon$ as well.

In three dimentional turbulence, since the evolution of $\epsilon$ is known to be slave to large scales, one expects that decay of dissipation, i.e., the right hand side of Eq. 2, to be independent of $Re_T$ in the limit of large Reynolds number. This implies that the divergent leading terms in the expansions of $Y$ and $P_\epsilon$ versus $Re_T$ must exactly cancel. This has been confirmed numerically as shown in Fig. 2, since the data of the combined terms is very well aligned with the horizontal black solid line. With this observation in mind, we aim to provide a model for the combined term $P_\epsilon - Y$. Additionally, in what follows, we still allow Reynolds-dependent model coefficients, but we use expansions that are convergent in the limit of large Reynolds numbers.

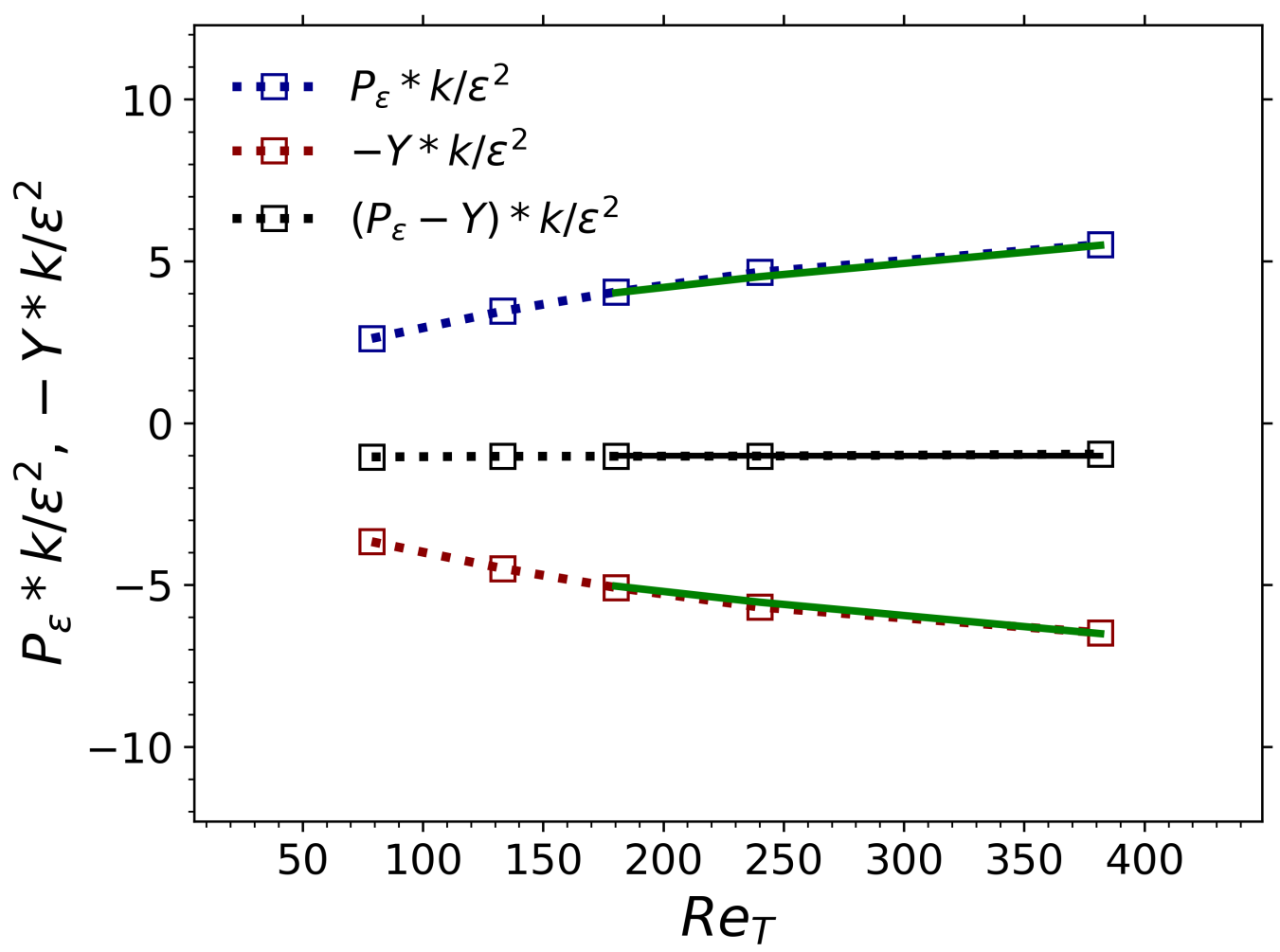

**Figure 2. Normalized $P_\epsilon(Re_T)$ and $-Y(Re_T)$ as a function of Reynolds number obtained from simulations of stationary turbulence. The overall normalized decay term $(P_\epsilon(Re_T)$ - $Y(Re_T))$ is additionally shown in black. The green lines are binomial fits to the data of $P_\epsilon(Re_T) * k/\epsilon^2$ and $-Y(Re_T) * k/\epsilon^2$ involving $y_1 Re_T^{1/2} + y_0$, where $y_1 = 0.24$ and $y_0 = 0.81$ for $P_\epsilon(Re_T) * k/\epsilon^2$. For $-Y(Re_T) * k/k/\epsilon^2$, $y_1 = -0.24$ and $y_0 = -1.81$ .**

To assess the dynamic behavior of the dissipation decay term, aside from analyzing stationary turbulence, we examine scenarios involving turbulence decay and growth. To initiate the decay simulation, an ensemble of initial conditions is sampled from the stationary simulation results. For each decay case, identified by its pre-decay Reynolds number, around 50 simulations are performed using such time-spaced initial states, creating ensemble-averaged data for $\epsilon$ and $P_\epsilon - Y$ from which the implied dynamic $C_{\epsilon_2}$ can be computed as a function of time. Due to the box size limit, each case was allowed to decay up to $t/T = 1.4$ to ensure that the turbulence correlation length remains reasonably small. Then, we used the final state of decay as the initial condition for the forced turbulence simulations with time constant forcing parameter $\Omega A_{ij} = \delta_{ij}$, allowing the turbulence to grow and reach about the initial value of TKE before the start of the decay process. Figure 3 shows $Re_T$ vs. time as turbulence decays and subsequently grows for each initial Reynolds number case.

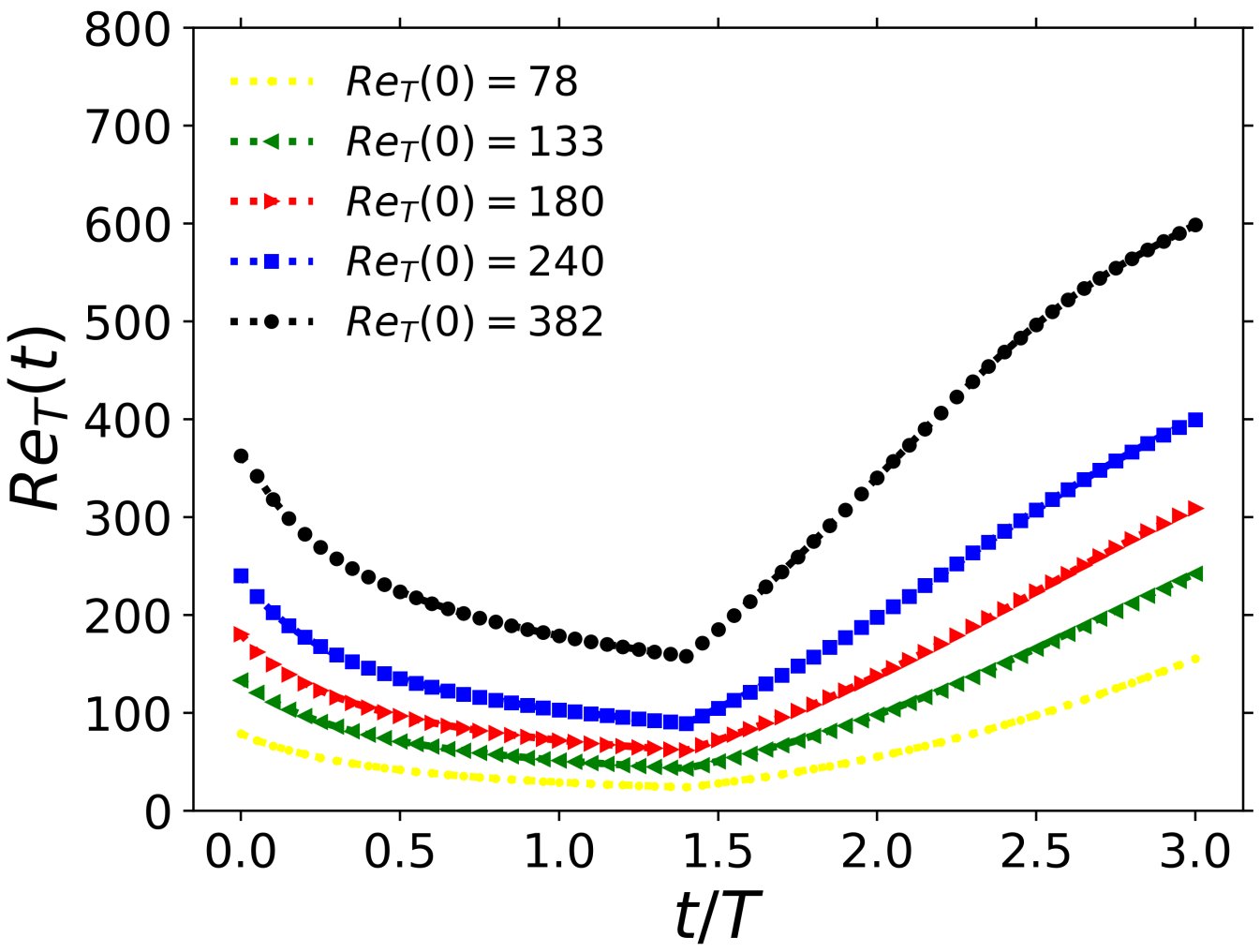

**Figure 3. Evolution of the Reynolds number during turbulence decay and growth. Each curve corresponds to a different case, distinguished by its initial Reynolds number.**

With these simulations, from the post-processing of data of TKE, $\epsilon$, and $P_\epsilon - Y$, one can estimate the implied $C_{\epsilon_2}$ inferred from combining Eqs. 6 and 7 as $C_{\epsilon_2} = -(P_\epsilon - Y) * k/\epsilon^2$. Figure 4 shows DNS results of computed $C_{\epsilon_2}$ for decay and forced turbulence. As shown in Fig. 4a, for the decay turbulence $C_{\epsilon_2}$ evolves in time, with a sharp increase from the initial value, then it exhibits a slower variation that remains Reynolds number dependent. For the growth of turbulence shown in Fig. 4b, the variation in time is evident for all cases. The values of $C_{\epsilon_2}$ reduce over time and drop below unity for all scenarios in the presence of turbulence production. Notably, higher Reynolds numbers demonstrate a more significant reduction in $C_{\epsilon_2}$ values before beginning to rise again.

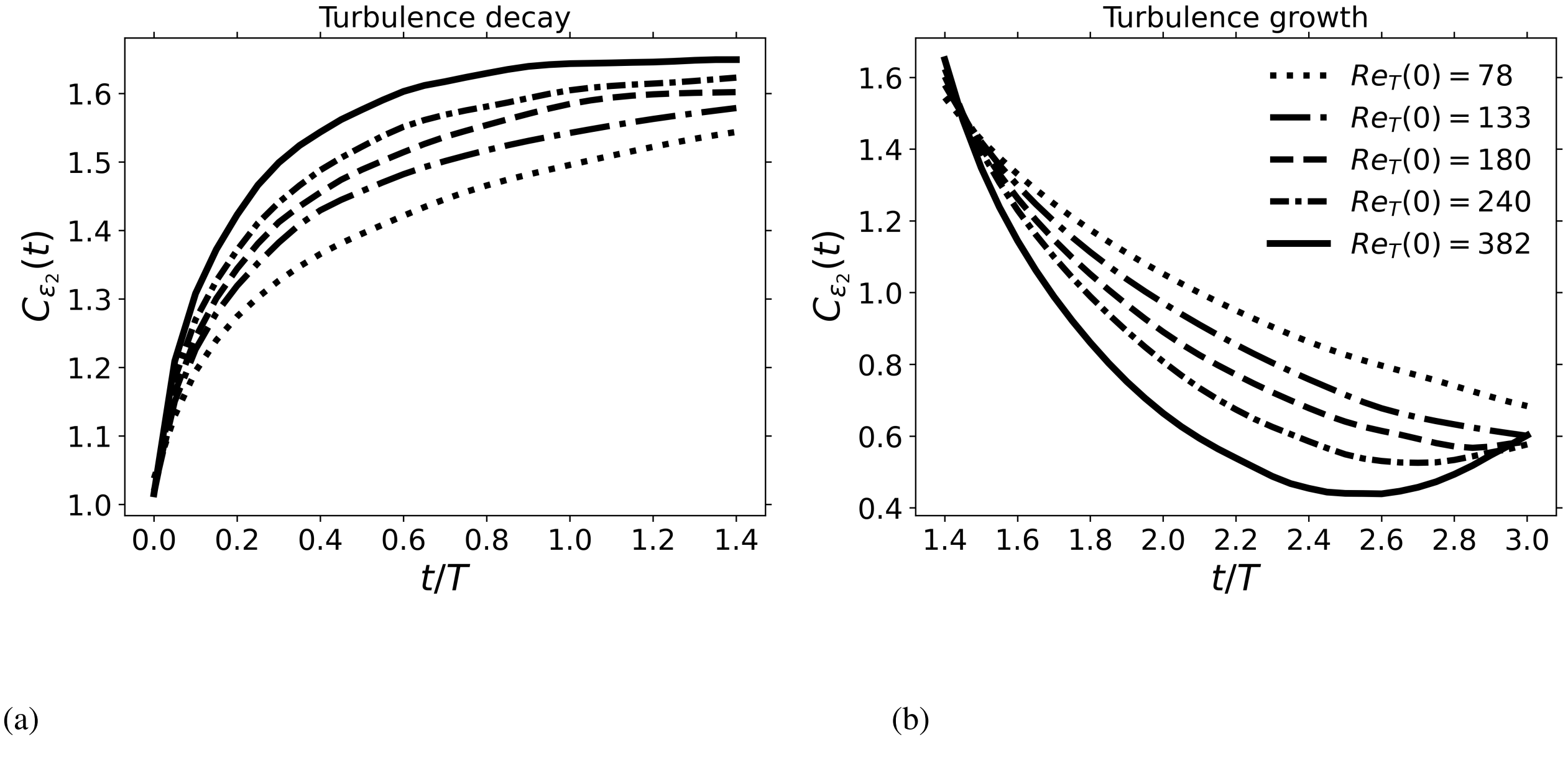

(a) (b)

**Figure 4. Evolution of $C_{\epsilon_2}$ in time for various initial Reynolds number values for (a) turbulence decay and (b) turbulence growth.**

Figure 5 shows $C_{\epsilon_2}$ vs. $Re_T$ for various cases. It is evident that each case is distinct, thereby cases are not collapsing. This is a confirmation that temporal dynamics, in addition to the local Reynolds number, play an important role in the evolution of $C_{\epsilon_2}$. Therefore, a model should incorporate these unsteady effects to accurately reproduce the dynamics of $C_{\epsilon_2}$. Based on these qualitative observations, we hypothesize that as turbulence decays, $C_{\epsilon_2}$ asymptotes to a value between 1.5 and 2 that is Reynolds number dependent. Conversely, when energy is injected in turbulence, via production, $C_{\epsilon_2}$ decreases.

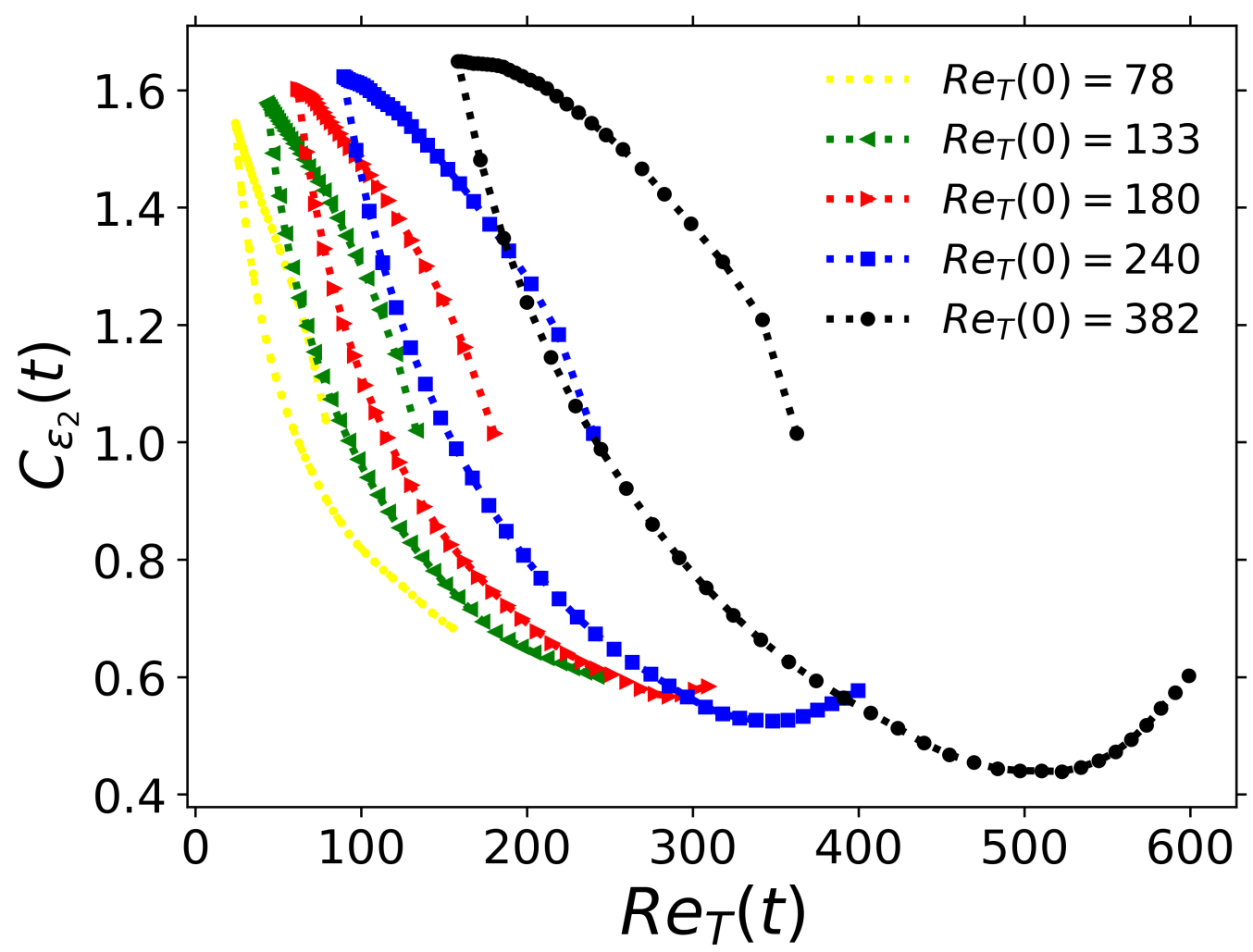

**Figure 5. Variation of vs. Reynolds number for the decay and growth of turbulence. Distinct graphs show various Reynolds numbers at the initial state.**

# 3 ANALYSIS AND MODEL DEVELOPMENT

Given the computed $C_{\epsilon_2}$, we developed a model for the evolution of $C_{\epsilon_2}$ that incorporates the effect of decay time and growth of turbulence as well as Reynolds number. Qualitative inspection of Fig. 4a indicates $C_{\epsilon_2}$ dynamics are similar to a first-order damped system. However, the target value towards which $C_{\epsilon_2}$ is being damped, and the damping rate may depend on the Reynolds number. Similarly, inspection of Fig.4b indicates changes in the target value when energy is injected in turbulence. Intuitively, the observed reduction in $C_{\epsilon_2}$ may be explained as follows. Without energy injection, the nonlinear cascade process takes energy from the large scale and distributes it to the small scale over a finite time. Under free decay, this process results in a certain distribution of kinetic energy across scales, which asymptotically leads to a specific value of $C_{\epsilon_2}$. However, energy injection due to the forcing term disrupts this process by injecting energy into all scales proportional to their kinetic energy. This energy injection not only changes the shape of the spectrum, but also creates shifts in the balance between $Y$ and $P_\epsilon$, which is otherwise present under free decay. This imbalance is particularly expected given that the dependence on the velocity is quadratic for the former term and cubic for the latter term. For example, in the extreme scenario of instant energy injection, at a time immediately after injection, the magnitude of $P_\epsilon$ is expected to increase more than $Y$, given this mismatch in the power law dependence of the two terms on velocity. We expect that in the case of continuous energy injection, considered here, the shift in the balance between $Y$ and $P_\epsilon$ not to be as severe as the case of instant energy injection, given the regularization effects of the cascade. However, given that both cascade and energy injection have comparable time scales, governed by the eddy turnover time, we expect a new steady state that is out of balance compared to that obtained in the case of free decay. As a result of this shift, we expect $Y$ to be less effective in overpowering $P_\epsilon$, and thus the overall decay term, $Y - P_\epsilon$, to be smaller compared to $\epsilon^2/K$ compared to the free decay scenario, leading to a smaller value of $C_{\epsilon_2}$.

Based on the above discussion, we expect the dynamics of $C_{\epsilon_2}$ to be sensitive to the nature of energy injection. The linear injection term considered in this study was selected to mimic the linearity of the shear production term seen in standard turbulence. Based on the observed trends in Fig. 4, we propose the following first-order model for the evolution of $C_{\epsilon_2}$.

$$\frac{dC_{\epsilon_2}}{dt} = A_D \frac{\epsilon - P}{k}(C_{\epsilon_D} - C_{\epsilon_2}) + A_F \frac{P}{k}(C_{\epsilon_F} - C_{\epsilon_2}), \tag{13}$$

where again $k$, $\epsilon$, and $P$ are respectively the turbulent kinetic energy, rate of kinetic energy dissipation, and rate of production. $A_D$, $A_F$, $C_{\epsilon_D}$, and $C_{\epsilon_F}$ are dimensionless parameters of this dynamical equation that must be determined and are expected to be Reynolds number dependent. In this notation, subscript $D$ is intended for parameters that capture the decay behavior, and subscript $F$ is intended for parameters that capture forced turbulence behavior.

To gain an intuitive understanding of this model, we first consider the case of decaying turbulence without production. Under this condition, the second term in the model will be zero, and we obtain a first-order system that damps $C_{\epsilon_2}$ towards a known $C_{\epsilon_D}$. The damping time scale is $k/\epsilon$ and its exact value is controlled by the dimensionless parameter $A_D$. In contrast, when the production term is active, the second term in Eq. 13 also becomes active, damping $C_{\epsilon_2}$ towards a new value. Specifically, in the case of stationary forced turbulence, where $P = \epsilon$, it is possible to reliably quantify the behavior by examining time-averaged results. Equation 13 is written such that the first term would vanish in this limit, while the second term would damp $C_{\epsilon_2}$ towards a new value, which we label as $C_{\epsilon_F}$.

Examining Fig. 4.b, it is worth noting that, depending on the Reynolds number, it is possible to observe non-monotonic evolution of $C_{\epsilon_2}$. As we shall see, the above model, once fully tuned, is capable of reproducing this nonmonotonicity as well as the aforementioned dynamics in the pure decay regime. Since we showed in Fig. 4 that the $C_{\epsilon_2}$ profiles are Reynolds number-dependent, in principle, all coefficients in Eq. 13 should also be Reynolds-dependent. Given the aforementioned analysis in Section 2.3, we anticipate that the leading order term capturing the Reynolds number effects will be proportional to the inverse square root of the Reynolds number in the high Reynolds flow limit. With this guiding principle in mind, and setting $C_{\epsilon_F}$ aside, we use the following functional forms for the remaining parameters.

$$A_D = a_{D_0} + a_{D_1}\left(a_{D_2} Re_T + 1\right)^{-1/2}, \tag{14a}$$

$$A_F = a_{F_0} + a_{F_1}\left(a_{F_2} Re_T + 1\right)^{-1/2}, \tag{14b}$$

$$C_{\epsilon_D} = C^{\infty}_{\epsilon_D} + (C^{0}_{\epsilon_D} - C^{\infty}_{\epsilon_D})\left(c Re_T + 1\right)^{-1/2}, \tag{14c}$$

where $a_{D_0}$, $a_{D_1}$, $a_{D_2}$, $a_{F_0}$, $a_{F_1}$, $a_{F_2}$, $C^{\infty}_{\epsilon_D}$ , $C^{0}_{\epsilon_D}$ and $c$ are constants to be determined. This functional form is written such that in the limit of $Re_T \to \infty$, the first coefficient represents the asymptotic value. The second terms capture the leading order finite Reynolds number correction, which are proportional to the square root of the Reynolds number in the high Reynolds number limit. A constant number one in the parentheses is added to all expressions to ensure a finite value of the parameters in the limit of zero Reynolds number. Nevertheless, this modification, if captured as a Taylor series expansion in the large $Re_T$ limit, still retains the inverse square root of the Reynolds number as the expansion parameter.

## 3.1 Model Tuning

**Table 3. Primary coefficients in Eqs. Eqs.13 and14c**

| $C_{\epsilon_F}$ | $C^{\infty}_{\epsilon_D}$ | $C^{0}_{\epsilon_D}$ |
|---|---|---|
| 1.00 | 1.73 | 1.50 |

We start this section by discussing the value of $C_{\epsilon_F}$. It is possible to show theoretically that under linear forcing $C_{\epsilon_F} = 1$ regardless of the Reynolds number. To do so, we consider the scenario of stationary forced turbulence in which production and dissipation balance each other. Under this condition Equation 13 reduces to

$$\frac{dC_{\epsilon_2}}{dt} = A_F \frac{P}{k}(C_{\epsilon_F} - C_{\epsilon_2}). \tag{15}$$

When flow is fully developed and reaches the stationary condition, the steady solution for this equation is $C_{\epsilon_2} = C_{\epsilon_F}$. Next, we examine the TKE and dissipation equations under this condition:

$$\frac{dk}{dt} = -\epsilon + 2Ak, \tag{16}$$

$$\frac{d\epsilon}{dt} = -C_{\epsilon_F}\frac{\epsilon^2}{k} + 2A\epsilon. \tag{17}$$

The second terms on the right-hand sides of these equations are production due to forcing, assuming an isotropic forcing tensor of the form $A_{ij} = A\delta_{ij}$ as discussed in section 2.1. Under the stationary forced condition, the left-hand sides of both equations vanish, and the two terms on the right-hand side must balance. Algebraic inspection reveals that a necessary condition of the simultaneous balance of both equations is $C_{\epsilon_F} = 1$ independent of the Reynolds number.

It is possible to measure $C_{\epsilon_F}$ by examining $C_{\epsilon_2}$ from the stationary forced simulations in our data sets and assess whether it is indeed equal to unity. Figure 6 shows the measured $C_{\epsilon_F}$ for the range of Reynolds numbers studied. One can see that the measured $C_{\epsilon_F}$ is close to the theoretical value of $C_{\epsilon_F} = 1$ within a small error and improves with increased Reynolds number. The figure quantifies the error associated with the statistical uncertainty of the measurements. The remaining gap between the theoretical value and the measured error is attributed to the fact that our forcing scheme uses the filtered velocity as opposed to the DNS field to avoid energy injection in scales that are comparable to the box size. Filtering this low-energy zone of the spectrum makes the second terms in Equations 15 and 16 slightly different from what we have theoretically assumed. However, for simplicity, we accept this error and proceed with $C_{\epsilon_F} = 1$. It is worth noting that in real scenarios where the production term is due to the mean velocity gradient tensor, as opposed to $A_{ij}$, all scales are forced by the mean velocity gradient without any high-pass filter.

Next, we discuss selection of $C^0_{\epsilon_D}$ and $C^\infty_{\epsilon_D}$ in Equation 14c. We remind that per our notation, $C_{\epsilon_D}$ represents the asymptotic value of $C_{\epsilon_2}$ in the long-term pure decay limit. Inspection of Equation 14c suggests that $C^0_{\epsilon_D}$ captures the regime of $Re_T \to 0$ and $C^\infty_{\epsilon_D}$ captures this values when $Re_T \to \infty$. Various analyses have studied the power law of turbulence decay in the low Reynolds number limit, under the so-called final period of decay [49, 50]. In these regimes, the role of inertia in the evolution of momentum fields is negligible, and the governing equations take a linear form. While significant research has focused on this regime, there is no consensus regarding the power law of decay of turbulence in this regime. We briefly provide a review of relevant literature below, before presenting our choice for $C^0_{\epsilon_D}$.

Early studies of the final period of decay for grid turbulence by Batchelor and Townsend concluded a temporal power law exponent of kinetic energy decay as $n = 2.5$ [50], resulting in $C^0_{\epsilon_D} = 1 + 1/n = 1.4$. A subsequent theoretical work by Saffman [51] resulted in $n = 1.5$, thus $C^0_{\epsilon_D} = 1.67$. This work assumed the low-wavenumber energy spectrum to follow $E(k) \propto k^2$. One can analytically derive the temporal exponent of $n = 1.5$ by assuming a viscous time scale for decay of each wavenumber, $\tau^{-1} \sim \nu k^2$, with active wavenumbers diminishing from the highest to the lowest. At each time instant, integrating the spectrum over the remaining wavenumbers, i.e., those that have not yet decayed, yields a kinetic energy decay exponent of 1.5. Same analysis with the assumption of low wavenumber spectrum scaling as $E(k) \propto k^4$, which is also called the Batchelor spectrum [52] would result in $n = 2.5$, consistent with the earlier experiments by Batchelor and Townsend [50]. Additionally, studies of isolated vortex rings [53] suggest that in their mature state, when the core size becomes comparable to the ring size, in the low Reynolds number limit, the vortex velocity scales as $t^{-3/2}$ and its length scales as $l \propto t^{1/2}$, which is also aligned with Saffman's analysis of vortex rings [54]. Modeling the turbulence as many isolated vortex rings, and accounting for local volumetric spreading of each ring, one can again arrive at $n = 1.5$ as the exponent for the decay of kinetic energy. The aforementioned studies [51, 53, 54], which suggest $n = 1.5$, i.e., $C^0_{\epsilon_D} = 1.67$, are either based on configurations that may not be relevant to grid turbulence, or idealized theories that involve simplifying assumptions regarding the dynamics of the energy spectrum.

Djenidi et al. [55] discussed that the realization of very low Reynolds number grid turbulence (i.e., the final stage of decaying turbulence (Batchelor and Townsend [50])) is challenging, as achieving homogeneous turbulence at low Reynolds numbers requires considerable downstream distances from the grid. Tan and Ling [49] offered an alternative interpretation of Batchelor and Townsend's data. By replotting the grid turbulence data, they found that the data tended to approach an inverse square, $n = 2$, energy decay law. Skrbek and Stalp [56] revisited the subject of decaying homogeneous isotropic turbulence by developing a phenomenological model based on different forms of three-dimensional energy spectra. Their model predicted a decay power law of $n = 2$, consistent with the experimental results of De Silva and Fernando [57]. Touil et al. [58] studied the decay of turbulence in a bounded domain using LES and DNS, applying a low-wavenumber cutoff in simulations of isotropic turbulence. In the final stage, they observed exponential viscous decay characterized by $n = 2$. In our study, we chose $C^0_{\epsilon_D} = 1.5$, which is compatible with $n = 2$ from these aforementioned computational and experimental results. This decision is made in light of our results presented in Fig.4a. It is apparent that for the low Reynolds limit, our asymptotic value of $C_{\epsilon 2}$ is below 1.67 ($n = 1.5$). Lower values such as $C^0_{\epsilon_D} = 1.5$ or 1.4 are more aligned with the trends observed in the figure. However, we avoid using $C^0_{\epsilon_D} = 1.4$, given the limited recent research in support of $n = 2.5$.

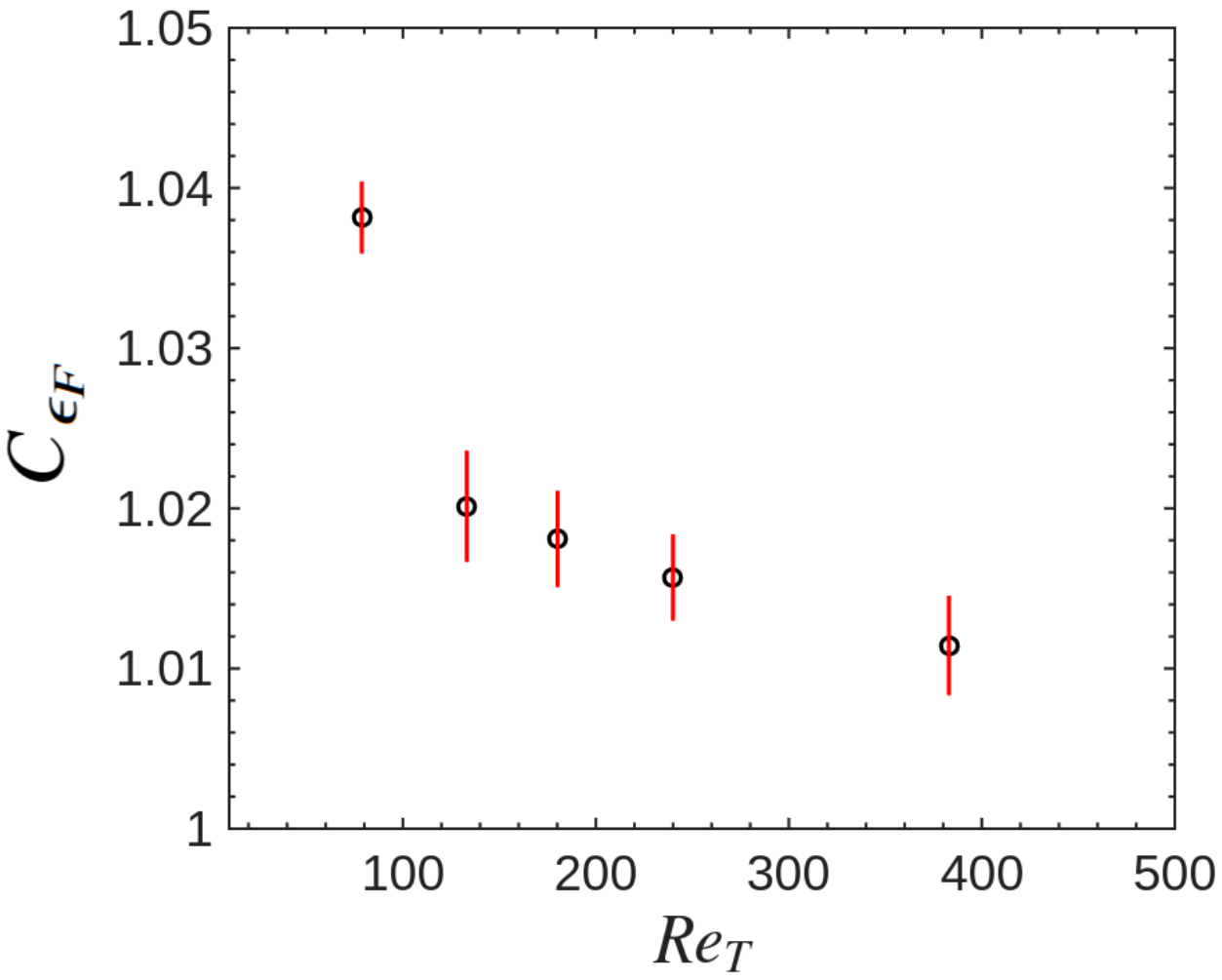

**Figure 6.** $C_{\epsilon_F}$ **vs Reynolds number for the stationary forced isotropic simulations. The vertical line indicates the** 95% **confidence interval associated with the statistical convergence error due to finite sample size.**

Next, we discuss the selection of $C^\infty_{\epsilon_D}$. The power-law decay exponent in the long-time, high-Reynolds-number limit, which is given by $1/(C^\infty_{\epsilon_D} - 1)$, has been studied analytically, experimentally, and computationally. Early analytical results indicate that the decay exponent $n$ is sensitive to the low-wavenumber (infrared) structure of the energy spectrum. In particular, for $E(k) \propto k^4$, corresponding to the Batchelor spectrum [52], Kolmogorov derived $n = 10/7$ [59]; for $E(k) \propto k^2$, corresponding to the Saffman spectrum [51], one obtains $n = 6/5$. As discussed in Appendix A, the classical $k^2$ and $k^4$ decay laws rely on the approximate persistence of the large-scale structure during decay, referred to as the permanence of large eddies (PLE) assumption [60, 61]. A common interpretation is that the infrared structures evolve on their own "eddy-turnover" timescale. However, this interpretation is questionable. In particular, the infrared spectrum may be influenced by the backscatter from the energetic eddies at higher wavenumbers, and thus evolve on a faster timescale. Appendix A presents our results on both the infrared spectrum scaling and the associated infrared timescale. The measured scaling differs from that predicted by the local eddy-turnover-time interpretation. This result weakens our confidence in selecting $C^\infty_{\epsilon_D}$ based on the aforementioned classical theories.

A number of wind-tunnel experiments have focused on homogeneous turbulence generated by grids. The original experiment of Comte-Bellot and Corrsin [25] found their best temporal power-law fit to the turbulent kinetic energy during the early part of decay to correspond to $n = 1.25$. Later experiments show slightly different values. Krogstad and Davidson [28], reported $n = 1.22$. They employed a multi-scale grid method to determine the $n$ value, accounting for uncertainties associated with the choice of virtual origin. This value is corrected from their original work, which uses a conventional grid and reported $n = 1.17$ [62]. However, most experiments lack the ideal flow conditions; for instance, generating turbulence at high Reynolds numbers is very difficult in grid turbulence [62]. In addition, the reported data fall within the transition period, when $n$ may be different from its asymptotic long-term value [30]. No universal decay exponent has been confirmed throughout extensive computational studies. DNS of decaying isotropic turbulence have tended to report decay exponents in a broad range [29–31]. This is in part due to the transition and finite Reynolds number effects, which are the subject of the present study. However, when it comes to the long time limit of large Reynolds number flows, a critical constraint in achieving this limit is the finite box effects. Referring to our own work, and as mentioned earlier, each case in Fig. 4a was allowed to decay up to $t/T = 1.4$ to ensure that the turbulence correlation length remains reasonably small, i.e., $L_I < 2\pi/10$. Figure 7 reports the estimated values of $C^{\infty}_{\epsilon_D}$, which is computed at the final time of Figure 4a. Again, assuming that the finite Reynolds correction scales as $Re_T^{-1/2}$ to the leading order, these values are reported versus the inverse square root of the Reynolds number measured locally in time for each curve. A linear fit to the data, reveals the value of $C^{\infty}_{\epsilon_D} = 1.73$ in the limit of $Re_T = \infty$. This value corresponds to $n = 1.37$. This value is within the range of the values reported in the literature, but it is substantially above the relatively popular value of $n = 1.2$, which is aligned with Saffman's theory and some of the wind tunnel experiments. A factor that distinguishes our analysis from prior computational work in determining $n$ (in the limit of long decay at infinite Reynolds number) is our systematic accounting for finite Reynolds effects and avoiding simulation regimes that can be influenced by finite box size effects. Nevertheless, future studies with larger computational boxes and higher Reynolds numbers are needed to establish confidence in this result and to offer potential corrections.

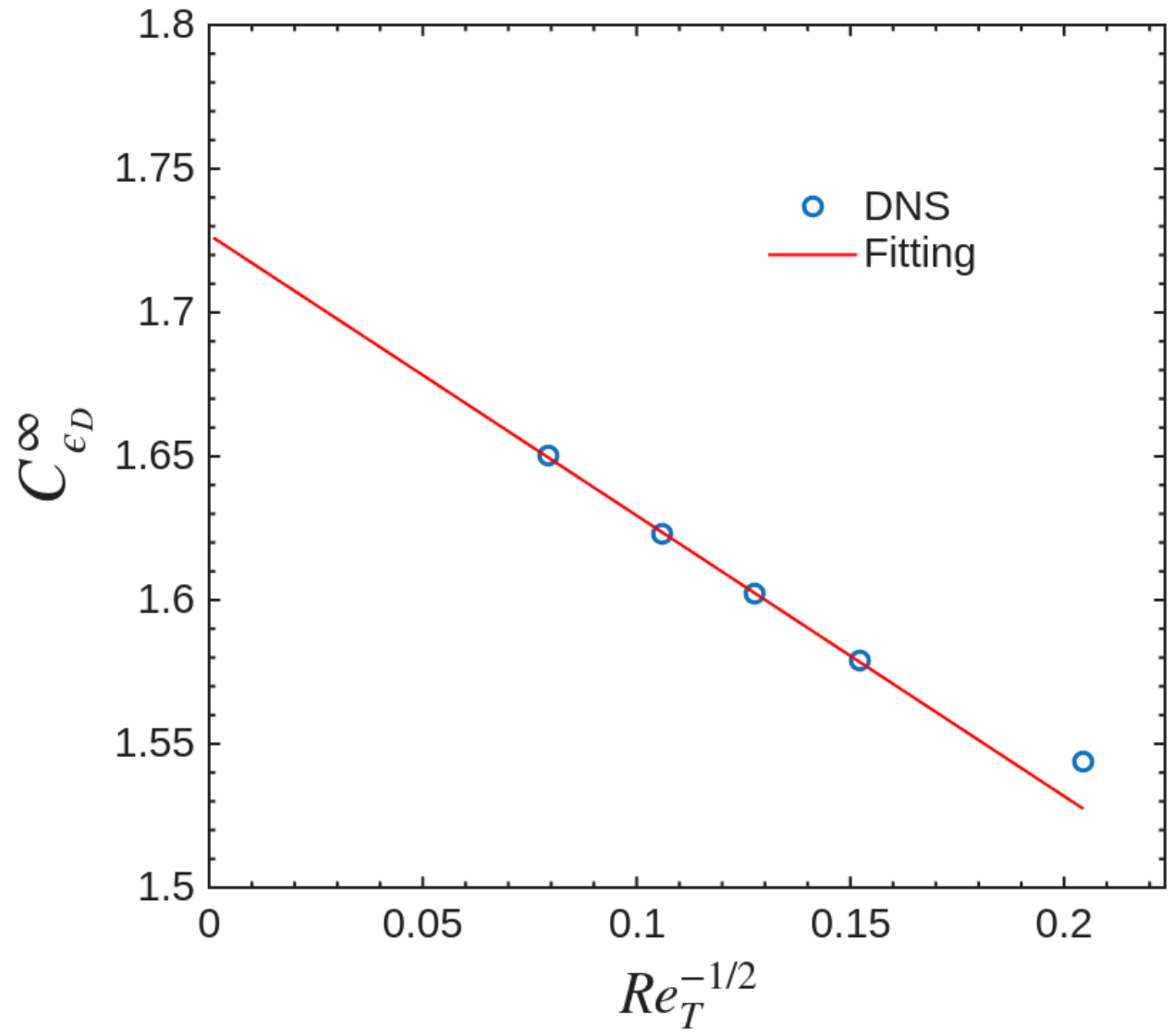

**Figure 7. $C^{\infty}_{\epsilon_D}$ versus Reynolds number, extracted from Fig. 4a at the final time. The red line represents a fit to the four cases with higher Reynolds numbers.**

Table 3 summarizes the primary coefficients for $C_{\epsilon_2}$ model (Eq. 14c), rounded to two significant digits. In order to tune the rest of the model's coefficients in Eqs. 14a, 14b,14c, we used data from our decaying turbulence simulations followed by forced turbulence at initial (pre-decay) Reynolds numbers $Re_T = 78$, $Re_T = 133$, $Re_T = 180$, and $Re_T = 382$. Data for $Re_T = 240$ was reserved to test the model. In our fitting optimization, we designed an outer iteration loop in which we searched for optimal values of $aF_{D_2}$, $a_{F_2}$ and $c$, while an inner loop was used to find $a_{D_0}$, $a_{D_1}$, $a_{F_0}$, and $a_{F_1}$ using least-square regressions given the linear dependence of the model on these parameters. The resulting rounded model coefficient values, which correspond to the lowest fitting errors, are provided in Table 4.

**Table 4. Resulting model coefficients for Eqs. 14a-c.**

| $a_{D_0}$ | $a_{D_1}$ | $a_{D_2}$ | $a_{F_0}$ | $a_{F_1}$ | $a_{F_2}$ | $c$ |
|---|---|---|---|---|---|---|
| 6.15 | -5.42 | 0.0032 | 3.17 | -3.36 | 0.058 | 0.21 |

An initial assessment of the tuned model against the same data set used for tuning is necessary to gain confidence regarding the developed model form and the selected coefficients. For the turbulence growth regime we solve Eqs.3-4, ($\frac{dk}{dt} = P - \epsilon$ and $\frac{d\epsilon}{dt} = P_\epsilon + F - Y$), where the decay term, $P_\epsilon - Y$, is modeled as $-C_{\epsilon_2}\frac{\epsilon^2}{k}$ along with Eq.13 for $C_{\epsilon_2}$ representing the RANS solution. Note that unlike the turbulence-decay regime, in which $F$ and $P$ are zero, in this case nonzero $P$ and $F$ are directly measured from DNS. Figures 8 and 9 show the evolutions of $C_{\epsilon_2}$, $\epsilon$ and $k$, for both decay and growth regimes respectively for Reynolds numbers, $Re_T = 180$ and $Re_T = 382$. The figures compared DNS against prediction of the developed RANS model. The figures also show a comparison to the standard $k-\epsilon$ model, which assumes $C_{\epsilon_2} = 1.92$. As these figures show, the model closely captures the evolution of quantities of interest, $k$ and $\epsilon$, for both decaying and forced turbulence.The variable $C_{\epsilon_2}$ model outperforms the standard $k-\epsilon$ model in predicting the nonlinear dynamics of decay and growth of TKE and dissipation.

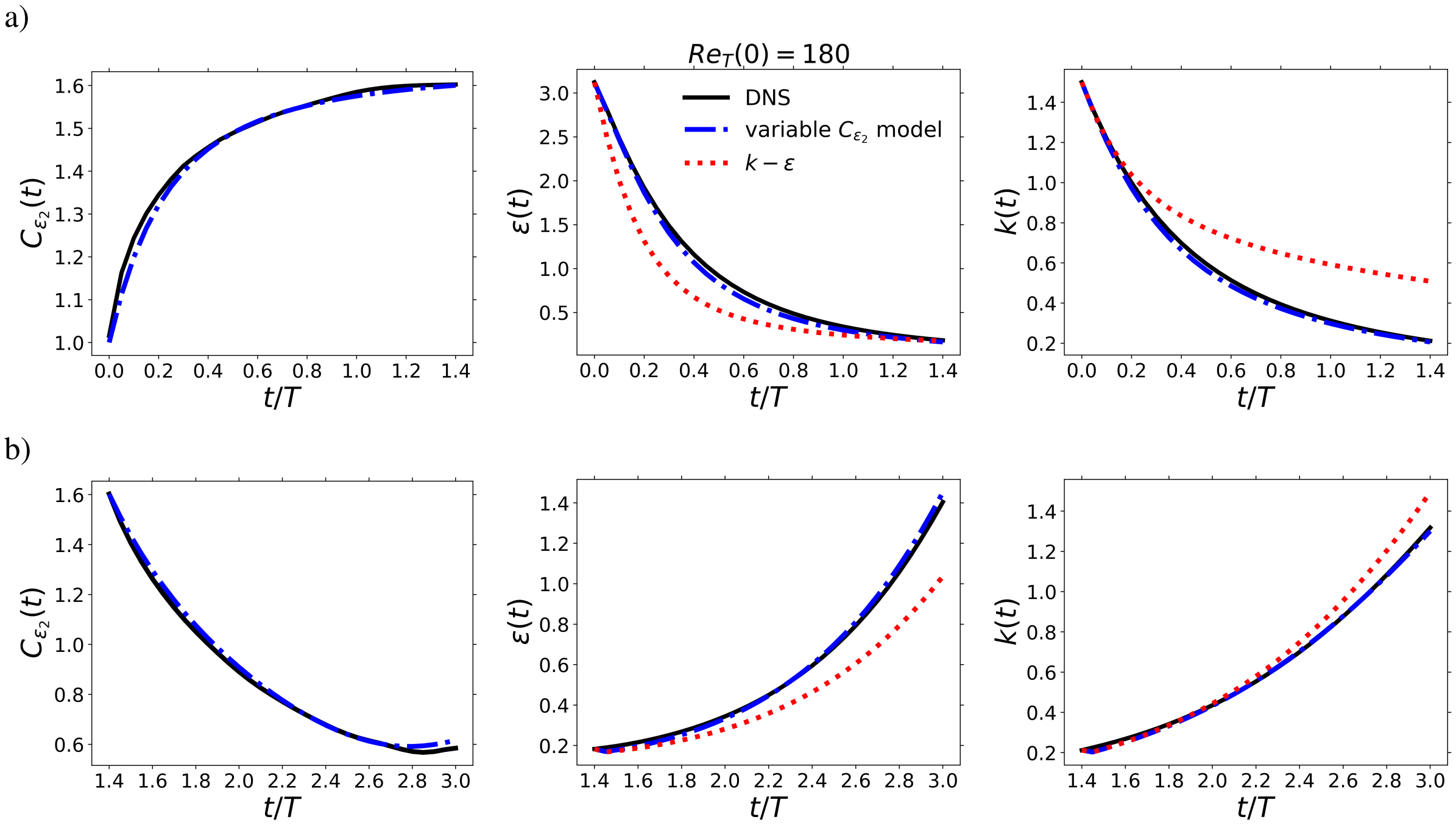

**Figure 8. Prediction of the developed model for evolution of decaying (a) and growing (b) turbulence compared to the DNS data and the standard $k-\epsilon$ model for the setting with the initial Reynolds number $Re_T = 180$.**

## 3.2 Model Assessment

To further assess the developed model, we next test it on a case not used for model tuning, namely $Re_T = 240$. Again, our RANS model solves for $k$, $\epsilon$, and $C_{\epsilon_2}$ using the developed coupled system of equations. Figure 10 shows the results of the model for turbulence decay and growth and in comparison against the DNS data as well as prediction of the standard $k-\epsilon$ model. For the decay case, the evolution of $C_{\epsilon_2}$ is well predicted. The prediction of $C_{\epsilon_2}$ for forced turbulence shows a reasonable match with the DNS data. Most importantly, incorporating the proposed evolutionary $C_{\epsilon_2}$ model into RANS, has resulted in an excellent prediction of the quantities of interest $k$ and $\epsilon$ in comparison to the reference data, offering substantial improvement over the standard $k-\epsilon$ model.

Next, we test the effectiveness of the model at an infinite Reynolds number, which is relevant to many practical turbulent flows. To approximate the infinite Reynolds number, we ran LES by specifying a zero value for the molecular viscosity. Following [4] we use LES in a box of size $2\pi^3$ with a mesh size of $64^3$, which was shown to result in a reasonable convergence in the turbulence statistics. In this case, following [4], post-processing of the terms in the turbulent kinetic energy evolution equation, and specifically the dissipation term, uses the variable eddy viscosity coefficient as $\epsilon = \overline{\nu_T \frac{\partial u_i}{\partial x_j}\frac{\partial u_i}{\partial x_j}}$, which is also equal to $-dk/dt$. The dissipation equation is not examined in this case, as dealiasing the triple products involving the variable $\nu_T$ would be cumbersome. Due to similar reasons, the LES results

a)

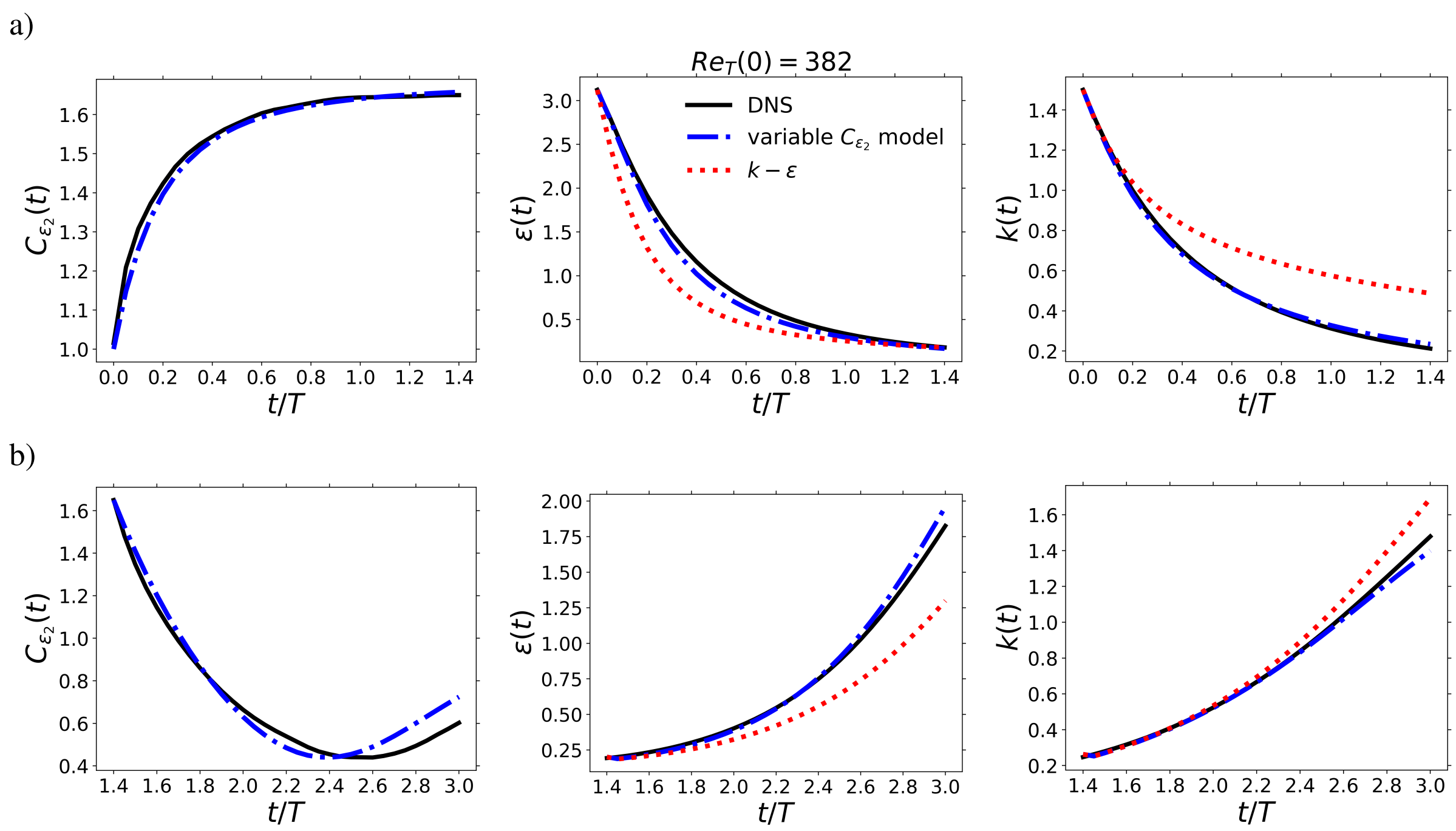

b)

Figure 9. Prediction of the developed model for evolution of decaying (a) and growing (b) turbulence compared to the DNS data and the standard $k - \epsilon$ model for the setting with the initial Reynolds number $Re_T = 382$.

a)

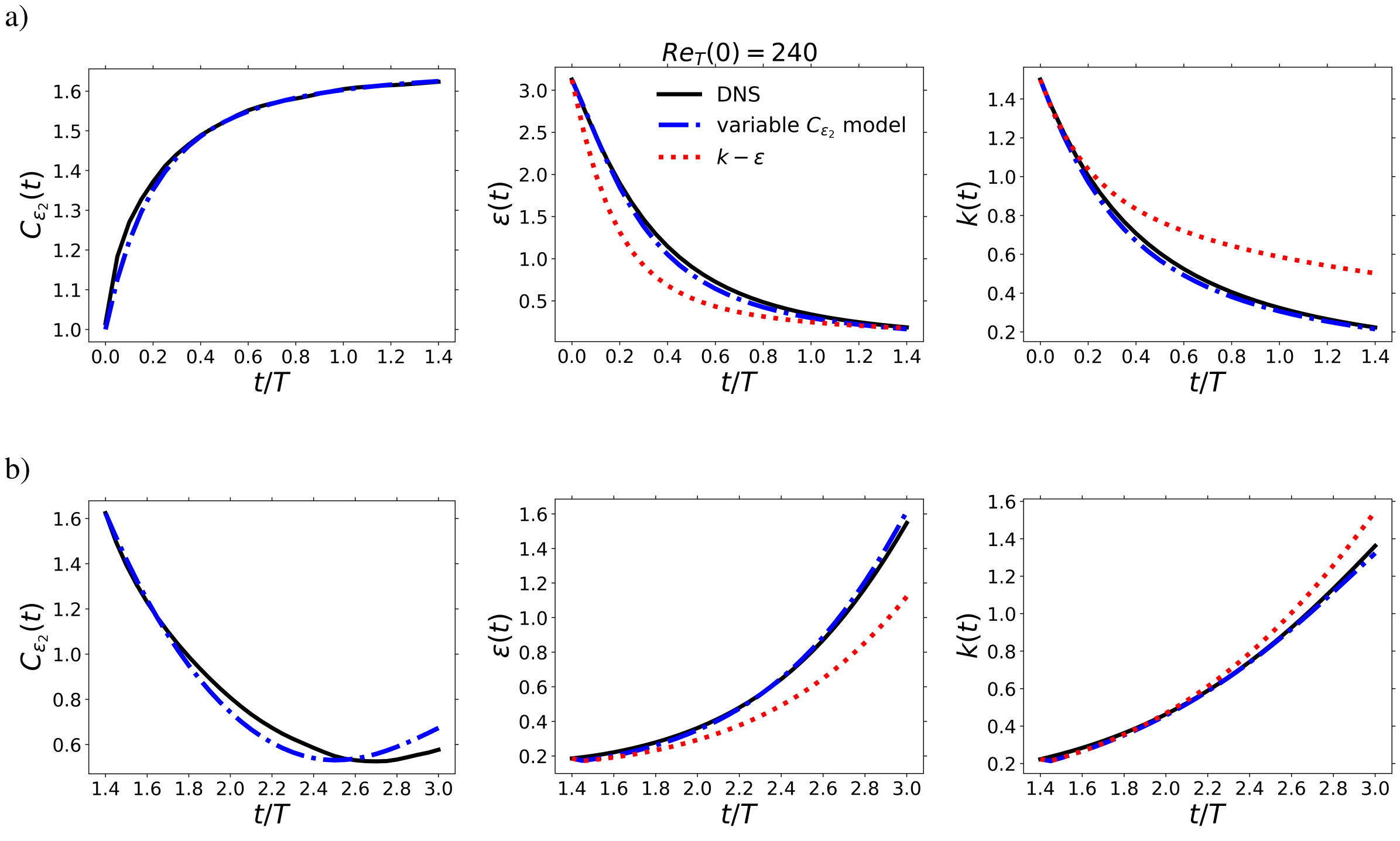

b)

Figure 10. Prediction of the developed model for evolution of decaying (a) and growing (b) turbulence compared to the DNS data and the standard $k - \epsilon$ model for the setting with the initial Reynolds number $Re_T = 240$.

are shown only for the decay case. For the cases involving turbulence growth via production, the production terms in the dissipation equation are dominated by small scales; therefore, it is challenging to recover them from LES data. As shown in Fig. 11b, the $k - \epsilon$ model does not predict the rapid decay of TKE, while invoking the variable coefficient model can largely correct the prediction of decay turbulence.

Finally, we test the model in the turbulence growth regime using energy injection rates different from those used for the generation of the tuning data. Specifically, we conducted new simulations of the growth regimes with coefficient $A$ at 50% less and 50% more than the nominal value, $A = 1$. This was done to demonstrate that the model performance is insensitive to the ratio of the turbulence growth time scale to its large eddy turnover time. Too small values of $A$ were avoided, since they would represent a regime close to the decay regime, which is already considered. Too large values would result in a fast increase in the Reynolds number and limit the time horizon over which the spatial mesh would be appropriate for the DNS solution. Figure 12a and b show the forced turbulence results at $A = 0.5$ and $A = 1.5$, respectively. For $A = 0.5$, dissipation decays initially. Then, after a period, which we attribute in part to the finite time of cascade for the injected energy to reach small scales, we observe dissipation growth. While the $k - \epsilon$ model also predicts the reversal in the evolution of dissipation, its predictions for both dissipation and kinetic energy are quantitatively inaccurate compared to the proposed model. For the higher energy injection rate, $A = 1.5$, due to the rapid energy growth, the mesh can be considered adequate for DNS up to $t/T = 2.4$. For this energy injection rate, the predictions of the $k - \epsilon$ model show discrepancies with DNS. In contrast, the variable $C_{\epsilon_2}$ model shows excellent agreement with DNS.

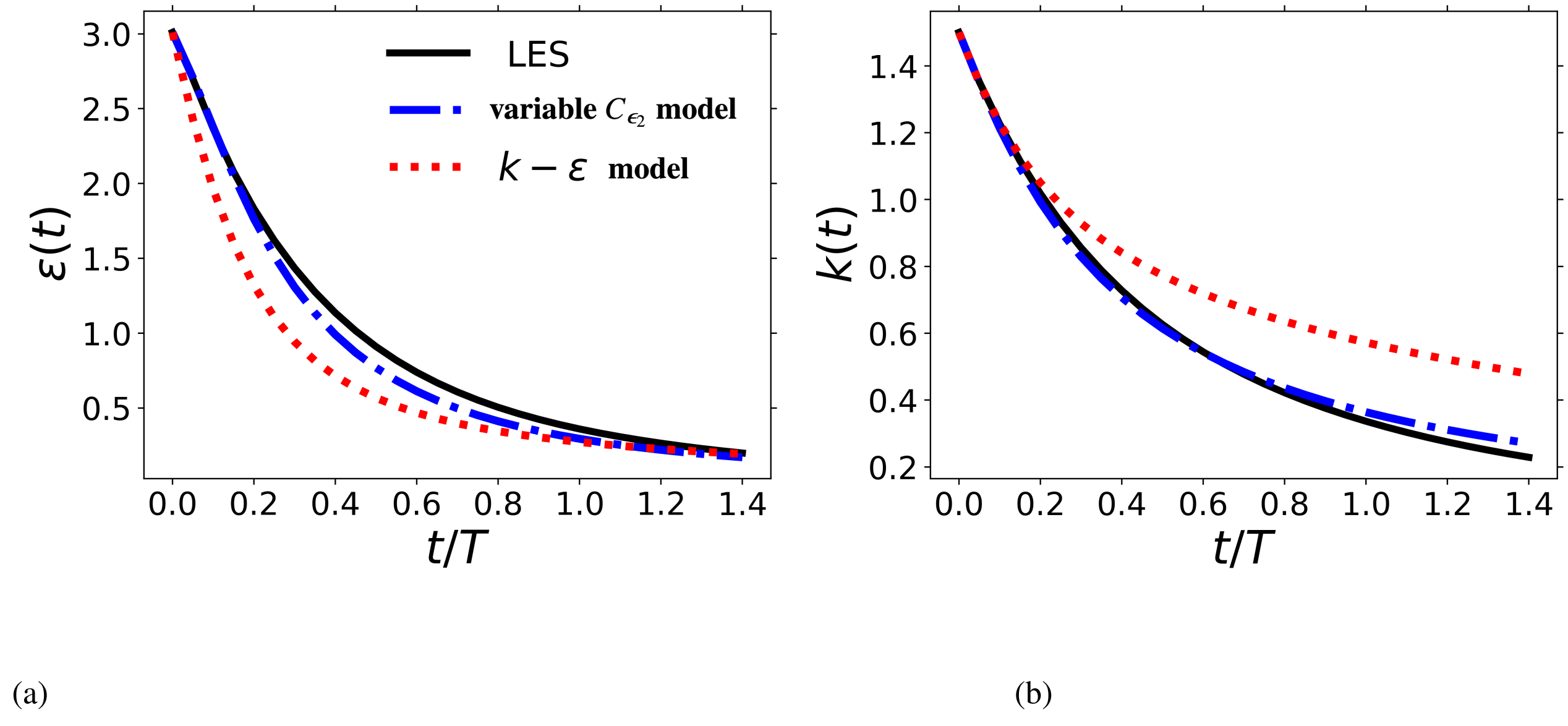

(a) (b)

**Figure 11. a) Evolution of dissipation and b) decay of TKE predicted by variable $C_\epsilon$ model compared to $k - \epsilon$ model at infinite Reynolds number.**

# 4 CONCLUSIONS AND DISCUSSIONS

This work is motivated by an understanding that, in two-equation turbulence models, an important parameter is the one determining the rate of homogeneous turbulence decay. This model parameter is important because the homogeneous decay term is always active. Additionally, this parameter is often tuned upfront, affecting the tuning of all subsequent parameters. In the $k - \epsilon$ modeling family, this parameter is referred to as $C_{\epsilon_2}$. The main product of this report is Eq. 13, which presents a time dynamic equation for the evolution of $C_{\epsilon_2}$, in the $k - \epsilon$ model family. At this point, this model is developed and tested for homogeneous and isotropic turbulence. Compared to a constant $C_{\epsilon_2}$, the proposed model captures the effects of local turbulence Reynolds number, as well as cascade dynamics with finite time scale.

The developed model is not derived; instead, it is proposed phenomenologically, based on observation of $C_{\epsilon_2}$

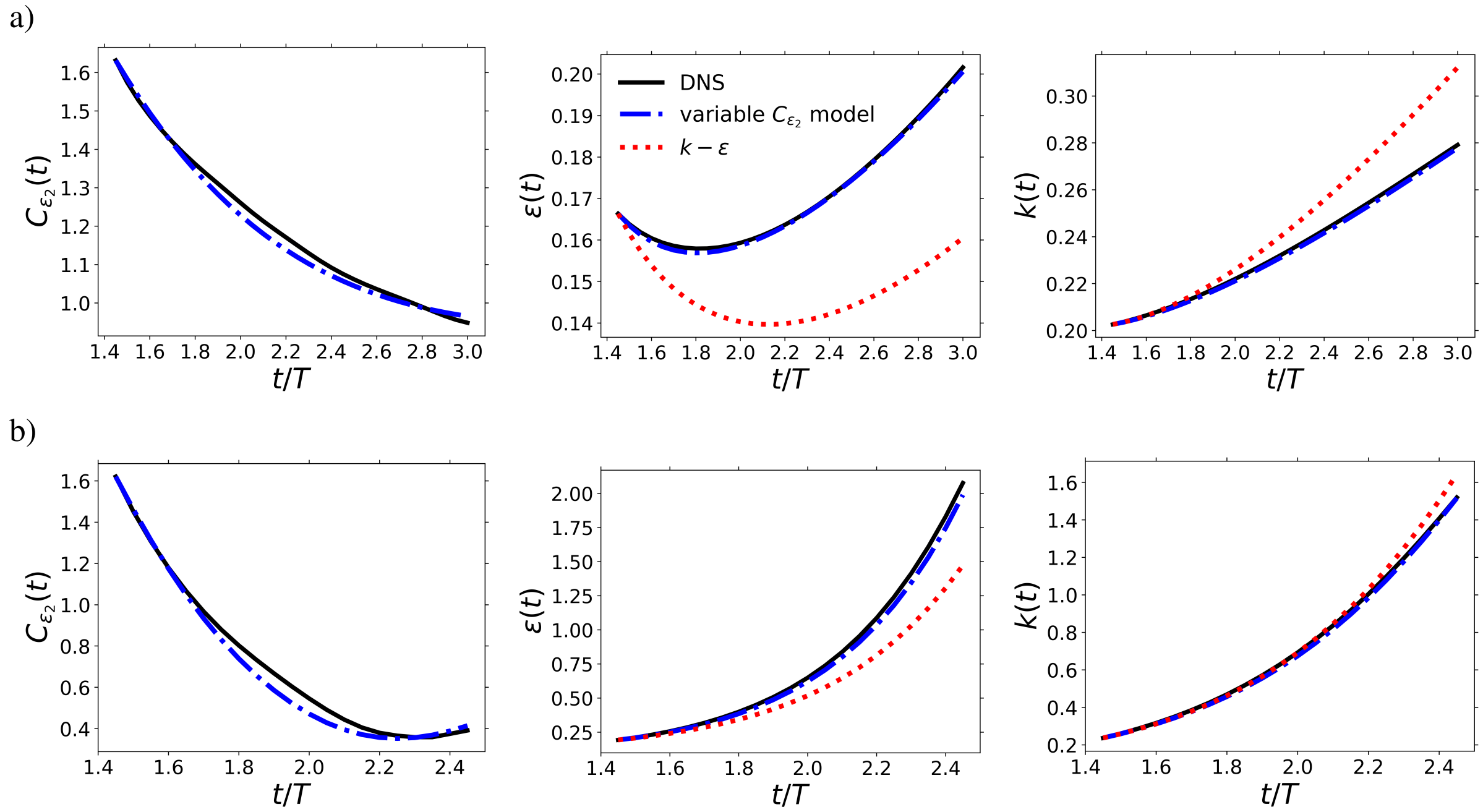

**Figure 12. Prediction of the turbulence growth regime using energy injection rates different from those used for the generation of the tuning data (a) at 50% less and (b) 50% more than the nominal value, $A = 1$ with the initial Reynolds number $Re_T = 240$.**

in a wide range of regimes in terms of Reynolds number, and turbulence time history involving a combination of free decay and active energy injection. The model form for the dependence on the Reynolds number was proposed based on a derived scaling analysis capturing the leading order effects to depend on the inverse square root of the Reynolds number.

We demonstrated the accuracy of the proposed model by testing the model in a RANS solver and showed that the model closely captures the evolution of quantities of interest, $\epsilon$ and $k$, for both decaying and growing turbulence and outperforms the standard $k-\epsilon$ model. This improved performance was shown to sensibly extend to regimes outside of those used for model tuning, including scenarios with different rates of energy injection compared to turbulence time, and an LES case representing an infinite Reynolds number.

While we consider this work as an incremental step towards improving existing turbulence models, multiple future steps should be considered before establishing full confidence in our results and before the utilization of the provided model form for CFD solutions. Some of such key items are provided below.

First, our data considered a limited range of Reynolds numbers and decay times, constrained by our turbulence box size limitations. To strengthen the confidence in our results, it would be useful to examine simulations at higher Reynolds numbers and much larger box sizes, thereby accessing a wider dynamical range for the evolution of $C_{\epsilon_2}$.

Second, the presented model lacks a validation test case involving practically relevant flow scenarios. One prerequisite is the development of a $C_{\epsilon_2}$ evolution equation, which also involves transport terms, extending its applicability to inhomogeneous flows. Additionally, one needs to go one step further and retune the entire $k-\epsilon$ model with the new $C_{\epsilon_2}$ model at hand. Although both of these steps are outside the scope of this work, they are the next natural step needed for extending this work to practically relevant flows.

Third, it is possible to doubt the applicability of data from periodic simulations for tuning models for practical flows. As noted in this work, it is established that decay dynamics are dominated by the largest structures in the flow. However, practical flows have a finite spatial span of turbulence, in which the largest structures do not have neighboring structures similar to them. Use of grid turbulence in experiments, or periodic boxes in simulations, for tuning decay dynamics is based on a hypothesis that repetition of similar large structures in a vicinity has a negligible influence on the closure model coefficient. Designing new canonical settings that challenge or test this implied hypothesis is needed before further establishing confidence in the practical utility of the developed model.

Fourth, the value of $C_{\epsilon_2}$ in the asymptotic limit of stationary forced turbulence, which is set as $C_{\epsilon_F} = 1$ in this work, is obtained assuming a linear forcing coefficient. This forcing injects energy across scales proportional to the energy of each wavenumber. Our investigation indicates that kinetic energy productions that distribute energy across scales different from the considered linear forcing imply different values of $C_{\epsilon_F}$. For example, stationary simulations with energy injection limited to large scales would result in a steady $C_{\epsilon_2} 1$. The choice of linear forcing used in this study is justified by the observation that physical production due to the mean velocity gradient is also a linear forcing affecting all scales proportional to their energy. Nevertheless, we show in Appendix B that even when turbulence is generated via injection at large scales, the subsequent decay dynamics after removal of the forcing can be reasonably captured by the model proposed in this work.

Lastly, we note that our model is intended and tested for regimes of finite production $P \leq O(\epsilon)$. In dimensionless form, our model offers corrections that are linear with respect to $P/\epsilon$, while it is possible to envision a nonlinear dependence on this parameter. As a result, our model in its current form is likely not suitable for dynamics with extreme or abrupt injections of kinetic energy. In limits with large $P/\epsilon$, Equation 13 may lead to nonphysical or unstable solutions. The use of quadratic or higher order corrections with respect to $P/\epsilon$ may be needed to provide physical accuracy and stability.

# Appendix A

In this appendix, we first summarize the classical derivation of the TKE decay exponent for homogeneous isotropic turbulence in the limit $Re \to \infty$, under the permanence-of-large-eddies (PLE) assumption [60, 61]. We then use our numerical data to examine a related but stronger assumption: that the low-wavenumber dynamics are governed by the "eddy-turnover" timescale.

The PLE assumption states that, during decay, the infrared (low-wavenumber) part of the energy spectrum retains the form $E(k) \sim Ak^N$ with a time-invariant $A$ when $k \to 0$, as illustrated in Figure 13. In the figure, $l$ denotes the characteristic length scale associated with energetic eddies, and $k_l \sim 1/l$ is the corresponding wavenumber.

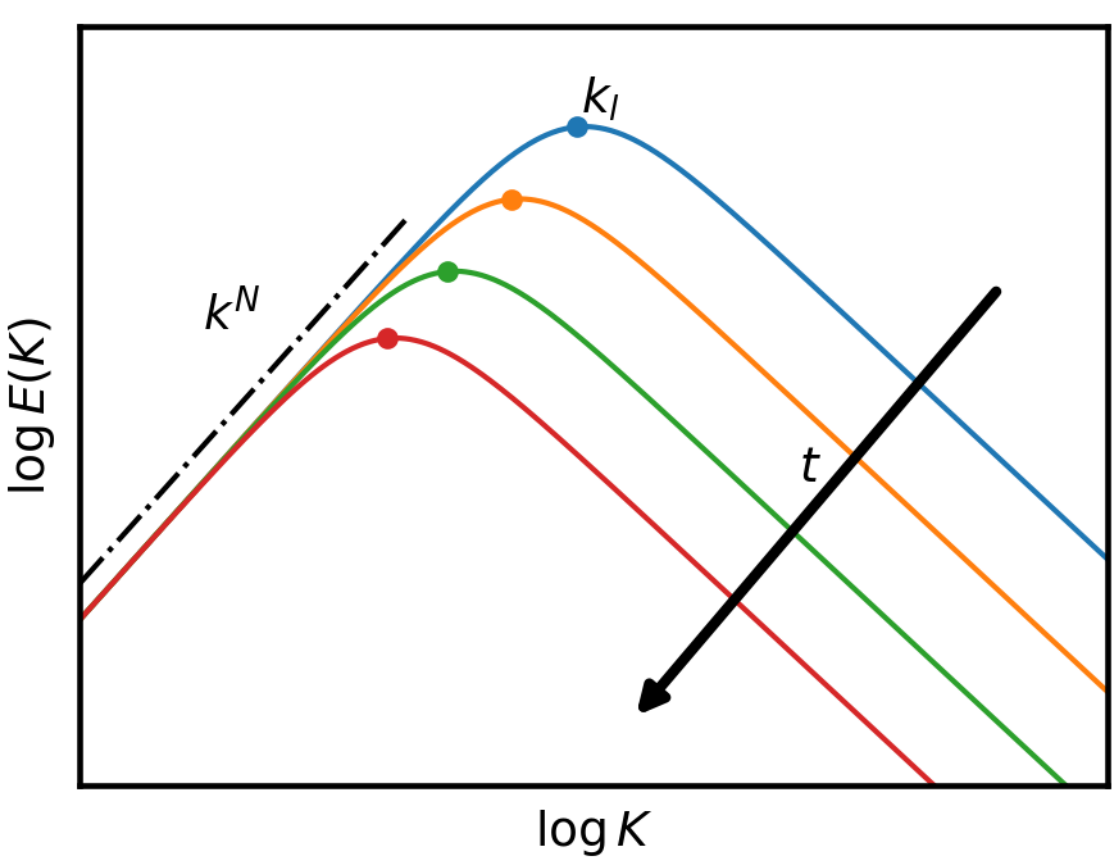

**Figure 13. Schematic illustration of the energy spectrum and its evolution during decay in the limit $Re \to \infty$. Under the permanence-of-large-eddies (PLE) assumption, the infrared spectrum retains the form $E(k) \sim Ak^N$ [60].**

The original derivations are given in [59, 63]. A summary is provided by [64]. Here, we restate the argument in our notation. The TKE scales as

$$u^2(t) \sim \int_0^{k_l} E(k)dk \sim Ak_l^{N+1}. \tag{18}$$

where $u$ indicates the RMS magnitude of velocity.

Following the zeroth-law of turbulence

$$\frac{du^2(t)}{dt} = -\epsilon \sim -u^3 k_l. \tag{19}$$

Eliminating $k_l$ using the infrared scaling above gives

$$\frac{du^2(t)}{dt} \sim -u^2(t)^{\frac{3N+5}{2(N+1)}}. \tag{20}$$

Assuming a power-law decay of the form $u^2(t) \sim t^{-n}$, substitution into the above relation yields

$$n = \frac{2(N+1)}{N+3}. \tag{21}$$

Therefore, for $N = 4$ we obtain $n = 10/7$ and for $N = 2$ we obtain $n = 6/5$. These are the classical Batchelor and Saffman decay exponents, respectively, as reported in the main text.

In the present study, the numerical results indicates that the infrared spectrum is of Saffman type. However, the numerical results do not show a persistent $n = 6/5$; instead, the decay exponent varies with time. To investigate this, we examine a related assumption to the PLE assumption, that the large scales evolve slowly on their own "eddy-turnover" timescale. As shown below, this assumption is not supported by our numerical results.

We first confirm the form of the infrared spectrum in our simulations. Figure 14 compares LES results obtained at the same resolution but with progressively larger domains, from a $2\pi$ box to an $8\pi$ box. The high-pass filter applied to the forcing is kept fixed in physical wavenumber among all cases, allowing forcings only with wavenumber $k > 2$. As the domain size increases, the infrared spectrum shows clear convergence toward $E(k) \sim k^2$, which is consistent with a Saffman-type spectrum.

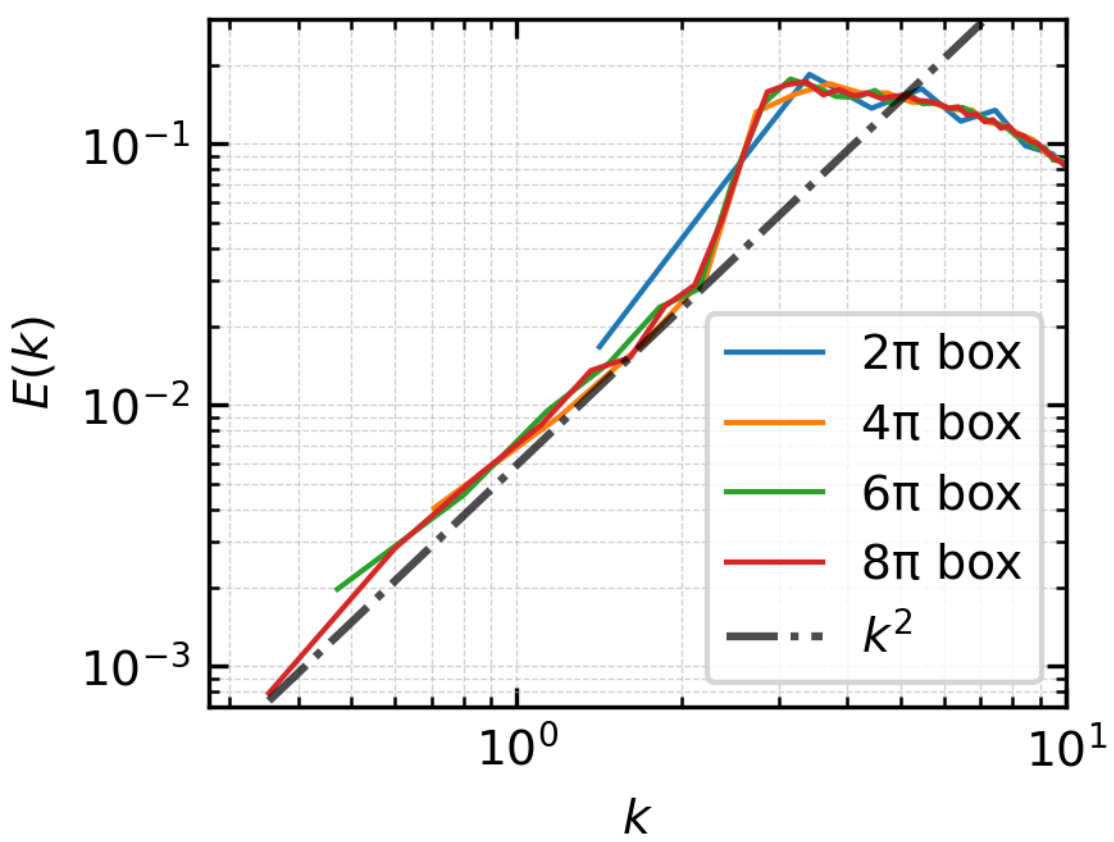

**Figure 14. Infrared energy spectra obtained from large-box LES. As the domain size increases, the infrared range approaches a $k^2$ scaling.**

We now turn to the assumption that the large scales evolve on their own "eddy-turnover" timescale. For a mode of wavenumber $k$, using $u_k^2 \sim AE(k)k \sim Ak^{N+1}$, the corresponding local turnover time is estimated as

$$\tau_k \sim \frac{l_k}{u_k} \sim \frac{1/k}{u_k} \sim A^{-1/2}k^{-(N+3)/2}. \tag{22}$$

For a Saffman spectrum, this gives $\tau_k \sim A^{-1/2}k^{-5/2}$.

We compare this scaling against the measurements from our simulations. Using the largest LES case ($8\pi$ box), we compute the temporal autocorrelation of the Fourier modes at fixed wavenumber $k$:

$$\rho_k(\tau) = C(k)\frac{1}{T_2 - T_1}\Re \sum_{\boldsymbol{k}:|\boldsymbol{k}|=k} \int_{T_1}^{T_2} \widehat{\boldsymbol{u}}^*_{\boldsymbol{k}}(t+\tau)\cdot \widehat{\boldsymbol{u}}_{\boldsymbol{k}}(t)dt, \tag{23}$$

where $\Re$ denotes the real part, $^*$ denotes complex conjugation, $\widehat{\boldsymbol{u}}_{\boldsymbol{k}}(t)$ is the complex Fourier coefficient of the velocity field $u$, and the averaging is taken over both time and all modes in the shell $|\boldsymbol{k}| = k$. The normalization constant $C(k)$ is chosen such that $\rho_k(0) = 1$. $T_1$ and $T_2$ represent the time window of the fully developed flow.

The autocorrelation is computed for modes with $0 < k \leq 2$ using time interval $T_1 = 150$ to $T_2 = 400$, including 2108 raw Fourier modes and 54 distinct shell-averaged wavenumbers. For each autocorrelation curve, the characteristic timescale $\tau_c$ is approximated from the time lag $\tau_{0.95}$ at which $\rho_k(\tau_c) = 0.95$ by $\tau_c = -\tau_{0.95}/\log(0.95)$. The corresponding results are shown in Figure 15. As expected, the autocorrelation decays more slowly for smaller $k$, but the measured timescale follows approximately $\tau_c \sim k^{-1}$ rather than $k^{-5/2}$, indicating that the infrared modes evolve faster than the "eddy-turnover" timescale, possibly due to backscattering from the energetic scales. This may provide a possible explanation for why the observed decay exponent does not follow the classical 6/5 value.

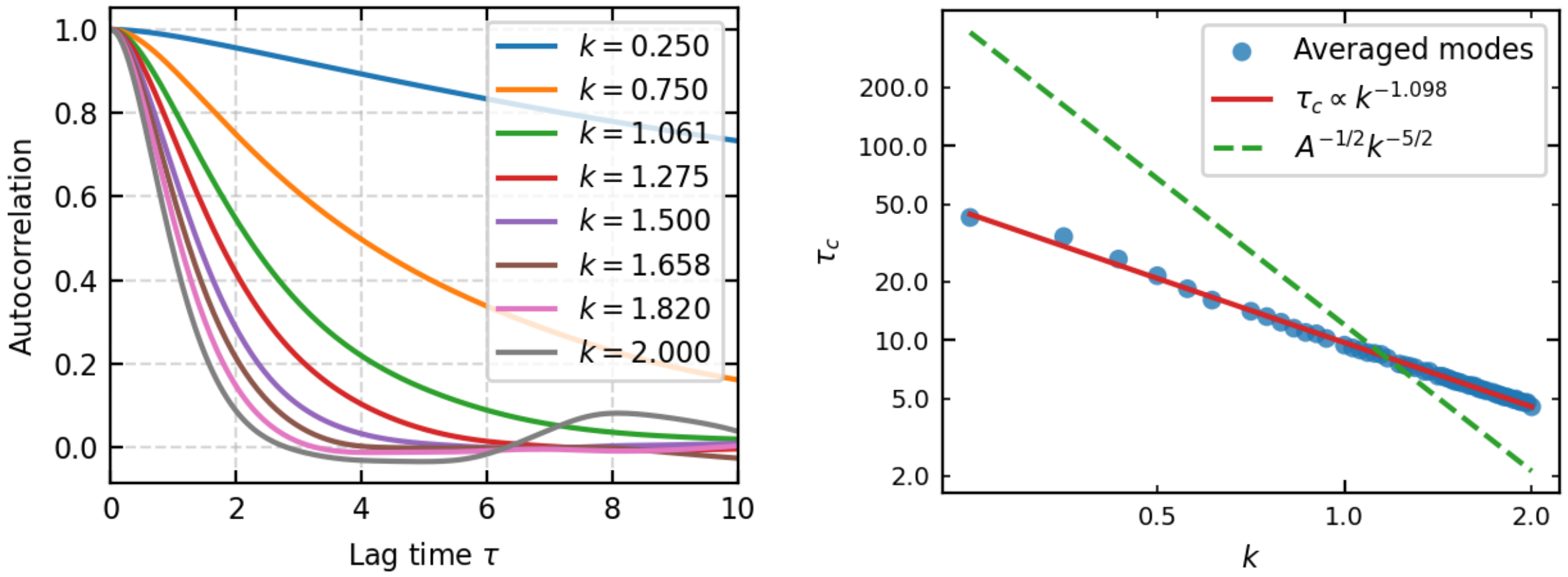

**Figure 15. Left: selected shell-averaged autocorrelation functions $\rho_k(\tau)$ for infrared modes. Right: characteristic timescale $\tau_c$ estimated from left panel, compared with the reference scaling $A^{-1/2}k^{-5/2}$, where $A$ is measured from Fig. 14.**

# Appendix B

In this appendix, we assess the proposed variable $C_{\epsilon_2}$ model against the data of decaying homogeneous isotropic turbulence when the initial turbulence is generated by purely large-scale forcing. This is in contrast to the linearly forced cases considered in the main text. The forcing is restricted to modes with wavenumber $2 \leq k \leq 4$. Two Reynolds numbers are considered, corresponding to the $N = 128$ and $N = 256$ entries in Table 1. For each case, the stationary forced turbulence is evolved for 400 time units. We then extract 50 snapshots of the 3D velocity field, separated by 4 time units, and use them as initial conditions for freely decaying simulations. The resulting statistics are ensemble averaged over the 50 simulations for each Reynolds number. The decay simulations are run up to 3 time units. Compared with the linearly forced cases, the present configuration reaches the domain-size limit later, allowing reliable statistics to be collected over a longer time interval.

Figure 16 compares the DNS-inferred $C_{\epsilon_2}$, the dissipation rate $\epsilon$, and the turbulent kinetic energy $k$ with predictions from the variable $C_{\epsilon_2}$ model and from the standard $k$-$\epsilon$ model. The comparison is started at $t = 1$ in order to exclude the initial transient immediately after the forcing is removed. The initial conditions for the RANS simulations are extracted from instantaneous and ensemble-averaged DNS data. The variable-coefficient model shows substantially better agreement with the DNS data for both resolutions.

One lesson from this exercise is the observation that for forced turbulence, $C_{\epsilon_2}$ is highly dependent on the distribution of energy injection across scales. Specifically, the data of $C_{\epsilon_2}$ at time $t = 0$ in the plots represent the fully developed forced conditions under our large-scale forcing. This value is substantially different from $C_{\epsilon_2} = 1$, which was shown for the case of linear forcing, i.e., injection at all scales proportional to their energy. Nevertheless, our comparison shows that once the turbulence is allowed to evolve away from this arbitrarily designed forcing, the dynamics of $C_{\epsilon_2}$ can be reasonably captured by the proposed model. Our interpretation is that, aside from extreme scenarios, the non-equilibrium dynamics of energy cascade across scales is likely governed by a low-dimensional system, at least for freely decaying homogeneous isotropic turbulence.

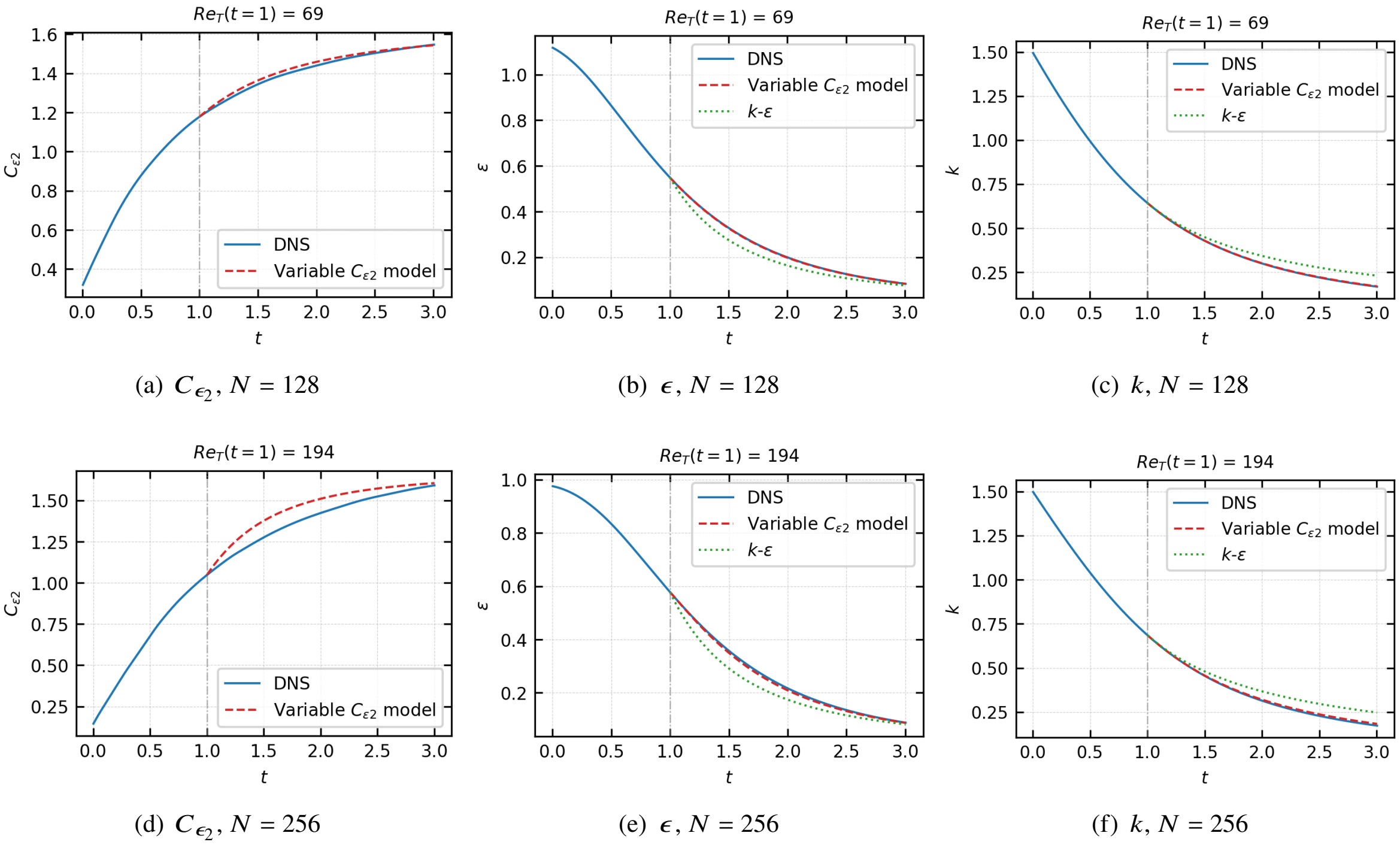

(a) $C_{\epsilon_2}$, $N = 128$ (b) $\epsilon$, $N = 128$ (c) $k$, $N = 128$

(d) $C_{\epsilon_2}$, $N = 256$ (e) $\epsilon$, $N = 256$ (f) $k$, $N = 256$

**Figure 16. Comparison of free-decay statistics initialized from stationary homogeneous isotropic turbulence sustained by purely large-scale forcing with $2 \leq k \leq 4$. Top row: $N = 128$. Bottom row: $N = 256$. Left to right: $C_{\epsilon_2}$, dissipation rate $\epsilon$, and turbulent kinetic energy $k$. In both cases, the variable $C_{\epsilon_2}$ model follows the DNS decay trends more closely than the standard $k$-$\epsilon$ model.**

# Acknowledgment

This work is supported by Boeing under Contract Number 2024-UI-PA-099 and NASA under Contract Number 80NSSC23M0225. The authors sincerely thank Dr. Chad Winkler for his valuable feedback and insightful reviews provided throughout this study.